\documentclass[iop,apj]{emulateapj} 
\usepackage{amsmath,amssymb,amstext}
\usepackage[breaklinks,colorlinks,citecolor=blue,linkcolor=magenta]{hyperref}

\usepackage[all]{hypcap}
\def\baselinestretch{0.96}
\usepackage{graphicx}
\usepackage{subfigure}
\usepackage{cleveref}
\usepackage{natbib}
\usepackage{epstopdf}
\usepackage{float}
\shorttitle{X-ray reprocessing in HMXBs}
\shortauthors{Aftab, Paul and Kretschmar}

\begin{document}

\title[]{X-ray reprocessing: Through the eclipse spectra of high mass X-ray binaries with \textit{XMM-Newton}}
\author{Nafisa Aftab$^{1}$}
\author{Biswajit Paul$^{1}$}
\author{Peter Kretschmar$^{2}$}
\email{Email of corresponding author: nafisa@rri.res.in, aftabnafisa@gmail.com}
\affil{{\small $^{1}$Raman Research Institute, C V Raman Avenue, Sadashivanagar, Bangalore 560080, India}}
\affil{{\small $^{2}$European Space Agency – European Space Astronomy Center (ESA-ESAC),
amino Bajo del Castillo, s/n., Urb. Villafranca del Castillo, 28692 Villanueva de la Canada, Madrid, Spain}}
%
\begin{abstract}
The study of X-ray reprocessing is one of the key diagnostic tools to probe the 
environment in X-ray binary systems. One difficult aspect of studying X-ray reprocessing is
the presence of much brighter primary radiation from the compact star together with the reprocessed radiation.
In contrast for eclipsing systems, the X-rays we receive during eclipse are only those produced by 
reprocessing of
the emission from the compact star by the surrounding medium. We report results from a
spectral study of the X-ray emission during eclipse and outside eclipse (when available) in
9 high mass X-ray binaries (HMXBs)
with  \textit{XMM-Newton} EPIC pn to investigate different aspects of the stellar wind in these HMXBs. 
 During eclipse the continuum component of the spectrum is reduced by a factor 
 of $\sim$8--237, but the count-rate for 6.4
keV Iron emission line or complex of Iron emission lines in HMXBs are reduced by a smaller
factor leading to large equivalent widths of the Iron emission lines. This indicates a
large size for the line emission region, comparable to or larger than the companion star
in these HMXB systems. However there are significant system to system differences. 
4U 1538$-$52, in spite of having
a large absorption column density, shows a soft emission component with comparable
flux during the eclipse and out-of-eclipse phases.
 Emission from Hydrogen-like Iron has been observed in LMC X-4 for the first time,
 in the out-of-eclipse phase in one of the observations. Overall, we find  significant differences
in the eclipse spectrum of different HMXBs and also in their eclipse spectra against  out-of-eclipse spectra.
\end{abstract}


\section{Introduction}

In X-ray binary systems, the X-rays produced very close to the compact object fuelled
by accretion are called primary X-rays. Usually a large part of these X-rays escapes directly
from the system and can be directly observed. But a significant fraction interacts with the 
matter surrounding the compact object and are re-emitted. This emission is called secondary emission.
Depending upon the interaction of the  primary X-rays with the surrounding
matter,  the secondary photons can have a wide range of energies.
This phenomenon  is known as reprocessing of X-rays (or X-ray reprocessing). The  
secondary emission as a whole is called reprocessed emission  and the secondary X-rays are
known as reprocessed X-rays. 
\par
Low energy primary X-rays colliding with very low energy electrons 
 through Thomson scattering are reflected with the same energy
 as they had before the interaction. 
Higher energy X-rays can interact with lower energy electrons,
giving off some of their energy to the electrons
via Compton scattering and coming out as lower
energy X-rays, UV, or optical photons. Moderate energy X-rays can be upscattered by relativistic electrons of the 
hot plasma and gain energy. Some X-rays interact with ionized or neutral atoms, excite electrons and give
rise to different spectral lines in the X-ray, UV, or optical bands. X-ray photons can be absorbed by photoionization 
of neutral or ionized atoms leading to reemission of photons at lower energies. 
Photons can also be affected by gravitational redshift. 
Reprocessing of X-rays is ubiquitous and has been observed in systems  
with accreting black holes, neutron stars, white dwarfs and in 
active galactic nuclei (AGN). 
\par
The X-ray reprocessing characteristics as seen from the Earth
depend upon several factors, for example 1) density and distribution of matter around the accretor;
2) structures in an accretion disc, if one is present 3) the orbital phase of the system; 
4) the chemical composition and ionization levels of the matter in the system;
5) the viewing angle of the observer.
Studying X-ray reprocessing has been proven to be a very useful tool to unveil the geometry,  
distribution of matter and its ionization level and also accretion mechanisms of accreting systems. 

\par
There have been theoretical and observational studies on X-ray reprocessing starting from  AGNs to low mass
X-ray binary (LMXB) systems, which have revealed different unique features of these systems, for example: 
AGN -- \cite{1994Zycki};
LMXB -- \cite{1996Jong}; Seyfert galaxy NGC 2992 -- \cite{1996Weaver}; supersoft X-ray source (SSS) CAL 87
(cataclysmic variable) -- \cite{2003Suleimanov}; black hole binary XTE J1817$-$330 -- \cite{2009Gierlinski} etc. 
\par
In this work we focus on the X-ray reprocessing in high mass X-ray binary (HMXB) systems. In
HMXB systems the companion star is a high mass star ($\geq$ 10M$_{\odot}$), either a main-sequence
star or a supergiant and the compact object is either a neutron star or a black hole. In HMXB systems
with a supergiant companion (SgHMXB system)  most of the time the compact object is embedded in the 
dense wind of the companion star, so the primary 
X-rays  encounter
scattering in these systems  which results into 
reprocessing.
Often the supergiant's wind contains clumps of different densities and sizes and intra-clump
 regions
are filled with rarefied wind materials \citep{oski2013}. 
The  reverse scenario is also possible, i.e., there could be lower density zones in a dense
 ambient medium.
This causes variation of X-ray reprocessing 
with
 the distribution of matter around the compact object. 
Supergiant fast X-ray transients (SFXTs) are a newer class of SgHMXBs 
\citep{2006Negueruela, 2005Sguera} which are usually dim sources
 (average luminosity 10$^{33}$-10$^{34}$ erg s$^{-1}$), but often appear with 
 unpredicted intense flares (flare luminosity reaches 10$^{36}$-10$^{37}$ erg s$^{-1}$).
However, most of the time the SFXTs have X-ray luminosity that is 3-4 orders of
 magnitude fainter compared to the classical HMXBs. 
 (luminosity at quiescent 10$^{32}$ erg s$^{-1}$). 
 These drastic change of luminosities in SFXTs indicates quite dynamic accretion scenario in these
 systems, hence variable wind structures around the compact object or different interaction
 between the compact object and companion's wind. In both the cases large variation in X-ray
 reprocessing is expected.
Accretion discs exist in
some HMXB systems (Cen X-3, LMC X-4, SMC X-1: \citealt{1978Savonije}; \citealt{2007Meer}).
Frequently structures (warps) form and decay in the accretion disc. The structures can differ in density,
temperature, shape etc. and face different
solid angle to the primary X-rays. Moreover these structures evolve with time, causing variation
in the X-ray reprocessing. 
\par
\par
Analyzing reprocessed X-ray emissions we can derive clues about the environment of the compact object in
an X-ray binary system. But the difficult aspect of studying reprocessed X-rays in X-ray 
binary systems is that the reprocessed emission is detected along with the primary 
emission from the compact object, 
which is much brighter. 
During eclipse the primary X-rays are blocked by the companion and the X-rays we receive
are the reprocessed X-rays only.
The intensity of the reprocessed X-rays is expected to be smaller
by a factor of a few. 
This increases the equivalent
width of a spectral line, when the emitting material is distributed in an extended region.
Thus the detection of a spectral line becomes more significant
during eclipse. These spectral lines give useful informations about
the chemical composition and ionization state of matter around the compact object.
When the compact object is not in eclipse, then the line intensity may be less significant
compared to the continuum and therefore difficult to measure accurately (if the line intensity
itself is not very intense). This results into low equivalent width and the line detection becomes
less significant during out-of-eclipse.
The comparison of  equivalent width during eclipse
can give useful insight about the distribution of different matter in different
binary systems and in a binary system at different epochs,  in much more efficient way 
compared to the out-of-eclipse phase. 
\par
Eclipse spectra of individual sources have been reported before 
(Cen X-3 -- \citealt{2012naik, 2001Wojdowski, 1996Ebisawa},
4U 1538$-$522 -- \citealt{2011Roca}, 4U 1700$-$377 -- \citealt{2005meer}, 
Vela X-1 -- \citealt{1999Sako,2002Schulz}).
The scattered reports on individual sources do not give an overall 
picture of the extent of
reprocessing during eclipse in HMXBs,
i.e, what is the fractional luminosity compared to
the out-of-eclipse spectrum, 
how do the shape of the continuum during the eclipse compare with the out-of-eclipse
spectrum, relative strength of the Iron (Fe) fluorescence line
 and other Fe lines if present. Moreover, we also aim to study system to system 
 variability, possible  
dependences on the type, mass, or mass loss rate of the companion and 
also to investigate the stability of the reprocessing in a given source when multiple data
sets exist. There indeed exist a large number of papers that investigate or use reprocessing as a
tool, but this is the
first work on a comprehensive analysis of all eclipse data 
available of HMXB systems with \textit{XMM-Newton} EPIC pn, an
 instrument particularly suitable
for its large collection area and CCD spectral resolution for the study of isolated emission lines.
\par
Out of 15 eclipsing HMXBs 9 sources have been observed with \textit{XMM-Newton}.
In this work we have analyzed all available observations of the
HMXBs with \textit{XMM-Newton} EPIC pn that include eclipses, i.e
in 7 Supergiant HMXBs or SgHMXBs (Cen X-3, 4U 1700$-$377, 4U 1538$-$52,
SMC X-1, LMC X-4, IGR J18027$-$2016, IGR J17252$-$3616) and in 2 
Supergiant Fast X-ray Transients
(SFXTs) namely IGR J16479$-$4514 and IGR J16418$-$4532.
We have focussed on these observations in order to have a coherent
sample of data for analysis and comparison.
A list of the important parameters of these sources are given in
Table  \ref{intro}. In the following paragraph we discuss some of the observed features of 
these systems which are relevant to our analysis. 
\begin{turnpage}
\begin{table*}[ht]
\begin{center}
\vspace{1ex}
\caption{List of basic parameters of the eclipsing HMXBs analyzed in this work, where $\mathrm{P_{orb}}$: 
Orbital Period (days), M$_{\rm C}$: Mass of the companion (M$_{\odot}$), $\mathrm{R_{C}}$: Radius of the companion star (R$_{\odot}$), 
$\mathrm{S_{C}}$: Spectral type of the companion,
e: Eccentricity of the HMXB orbit, a: Length of the semi major axis of the system (a$_{x}$sini [light-sec]) , 
${\dot{\rm M}}_{w}$ : Mass loss of the companion star (10$^{-7}$~M$_{\odot} \rm ~yr^{-1}$),  
d: Distance from Earth (kpc). (M$_{\odot}$, R$_{\odot}$: Mass and radius of Sun respectively)}
\begin{small}
\begin{tabular}{ll ll ll ll ll}
\hline
&&&&&&&&\\

  Source	&Mode of 	                   & $\mathrm{P_{orb}}$	& M$_{\rm C}$  	       & R$_{\rm C}$	      &	S$_{\rm C}$	  & e		           & a		    		& ${\dot{\rm M}}_{w}$              & d       \\
	        &accretion   &&&&&&\\

		\hline
			     &&&&&&&&\\

 Cen X-3 	&wind+disc$^{a1}$	   & 2.09$^{a2}$	&(20.5$\pm$0.7)$^{a3}$ & 12.1$\pm$ 0.5$^{a4}$ &O6.5 II-III$^{a4}$ &$\leq$0.0016$^{a5}$    & 39.6612$\pm$0.0009$^{a6}$   &  5.3$^{a1}$  & 5.7$\pm$1.5$^{a7}$  \\

			     &&&&&&&&\\
			
 LMC X-4 	&wind+disc$^{a1}$		        & 1.41$^{b2}$	& 14.5$^{+1.1}_{-1.0}$$^{a4}$	& 7.8$^{+0.3}_{-0.4}$$^{a4}$  &O8 III$^{a4}$	  &0.0006$\pm$0.0002$^{b3}$	   &26.343$\pm$0.016$^{b3}$ & 2.4$^{a1}$ &49.97$\pm$0.19$\pm$1.11$^{b4}$\\
 
					      		&&		&			&		&		&			&			& &\\
			
 SMC X-1 	&wind+disc$^{a1}$		      &3.89$^{c2}$		&(16.6 $\pm$ 0.4)$^{c3}$			& 18$^{c4}$		&B0 I$^{c5}$		&$<$0.0007$^{c4}$ &53.4876 lt-sec$^{c6}$   &	15$^{a1}$	&60.6$\pm$1.0$\pm$2.8$^{c7}$\\
 			&		      &			&			&		&		&			&			&& \\

 4U 1700$-$377	&wind$^{a1}$	   &3. 412$^{d1}$ 	&(52 $\pm$ 2)$^{d2}$ 	& 21.9$^{+1.3}_{-0.5}$$^{d2}$	&O7f$^{d3}$ 	& $<$0.008$^{d4}$			&48-82$^{d1}$ 			&$>$21$^{a1}$	&1.8$^{d5}$\\
  			&		      &			&			&		&		&			&		&& \\

 4U 1538$-$522	&wind$^{a1}$	          &3.75$^{e2}$     	&(19.9$\pm$3.4)$^{e3}$	& 17.2$\pm$1.0$^{e3}$	& B0Iab$^{e4}$	&0.174$\pm$0.015$^{e5}$		&  53.1$\pm$1.5$^{e5}$ & 8.3$^{a1}$ 	&6.4$\pm$1.0$^{e5}$\\

 			&		      &			&			&		&		&			&			& &\\
 
 IGR J18027$-$2016	&  wind$^{a1}$          & 4.56$^{f2}$		& 18.8--29.3$^{f2}$		& (15.0--23.4)$^{f2}$	& B1-Ib$^{f3}$		&6.3$^{f1b}$ 		& 68$\pm$1$^{f2}$ &6.3$^{a1}$	&-\\
 
		        &      		&&&&&&&&\\

 IGR J17252$-$3616  	&wind$^{a1}$	      &9.74$^{i2}$ &15$^{i3}$ 	&21--37$^{i2}$	& B0 I--B5 I$^{i2}$ 	&$<$0.19$^{i2}$ &101$\pm$3$^{i2}$  		&9.0 $^{a1}$  &5.3-8.7$^{i2}$\\
 					      
		        &      		&&&&&&&&\\

 IGR J16479$-$4514 	& SFXT$^{g0}$$^{,}$ $^{g1}$      & 3.32$^{g2}$ 		& 30$^{g3}$	& 23.8$^{g3}$	&O9.5 Iab$^{g4}$		&- 			& (44.85-48.90)$^{z1}$			 &(10-70)$^{y2}$       &2.8 $^{+4.9}_{-1.7}$$^{g4}$\\
  			&		      &			&			&		&		&			&			&& \\

 IGR J16418$-$4532 	& SFXT$^{h0}$     & 3.75$^{h2}$ 		& 31.54$^{h3}$ 		& 21.41$^{h3}$		& BN 0.5 Ia$^{h4}$ 	& -		& (48.65-53.03)$^{z2}$			&	(2.3-3.8)$^{y3}$ &13$^{h5}$\\
  			&		      &			&			&		&		&			&			&& \\

\hline
			&                     &       		&          		&		&  	& 		        &        			&  &       \\

 \end{tabular}
 \end{small}
\label{intro}
\\
   \footnotesize{$^{a1}$\citep{2015Falanga}, $^{a2}$\citep{1972SchreierL79}, $^{a3}$\citep{1999Ash}, $^{a4}$\citep{2007Meer},  
   $^{a5}$\citep{1997Bildsten}, $^{a6}$\citep{2010Rai}, $^{a7}$\citep{2009Thompson}, \\
   $^{b2}$\citep{1978Li}; \citep{1978White}, $^{b3}$\citep{2000Levine541}, $^{b4}$\citep{2013Pietrzy},  \\
   $^{c2}$\citep{1972Schreier}, $^{c3}$\citep{2005Baker}, $^{c4}$\citep{1977Primini}, $^{c5}$\citep{1993Reynolds}, 
   $^{c6}$\citep{1993Levine}, $^{c7}$\citep{2005Hilditch}     \\
   $^{d1}$\citep{1973Jones},  $^{d2}$\citep{2002Clark}, $^{d3}$\citep{1973Penny}, $^{d4}$\citep{2016Islam},  $^{d5}$\citep{2001Ankay} \\
   $^{e2}$\citep{1977Becker}, $^{e3}$\citep{1992Reynolds}, $^{e4}$\citep{1978Parkes}, \citep{2015Falanga}, $^{e5}$\citep{2006Uddi}, \\
   $^{f1b}$\citep{2003Augello}, $^{f2}$\citep{2005Hill}, \citep{2009Jaindec}, $^{f3}$\citep{2010Torrej}, \\   
   $^{g0}$\citep{2005Sguera}, $^{g1}$\citep{2003Molkov}, $^{g2}$\citep{2009Jain}, $^{g3}$\citep{1996Vacca}, $^{g4}$\citep{2008Nespoli},   \\   
   $^{h0}$\citep{2011Romano}, $^{h2}$\citep{2006Corbet}, $^{h3}$\citep{2005Martins}, $^{h4}$\citep{2013Coleiro},   \\
   $^{i2}$\citep{2007Thompson}, $^{i3}$\citep{1990Takeuchi} \\
   $^{y2}$\citep{2012sidoli} , $^{y3}$\citep{2013drave}  \\
   $^{z1}$, $^{z2}$  Derived from total mass of the binary and the orbital period, taken from \cite{2015Coley}} 
\end{center}
\end{table*} 
\end{turnpage}
\par
Cen X-3 was the first discovered X-ray binary pulsar  \citep{1971Giacconi} which 
accretes matter directly from the supergiant's wind and also through  disk accretion via 
Roche Lobe overflow \citep{1992Nagase, 1993Day}.
\cite{1992Nagase}
 observed an Fe line in Cen X-3 with energy $\sim$6.5 keV.
The intensity of this line was observed to be pulsating with phase, which indicates
that the matter  surrounding the neutron star causing this fluorescent line 
is not uniformly distributed \citep{1993Day}. 
\cite{1996Ebisawa} with 1993 ASCA observation have clearly resolved three 
Fe K$_{\alpha}$ emission lines with energy of 6.4 keV, 6.7 keV and 6.97 keV during eclipse.
The highly ionized 6.7 keV and 6.97 keV emission lines could not be resolved in
the pre eclipse phase. The line parameters indicate that the origin of the fluorescent
6.4 keV was nearer to the neutron star as compared to that of the highly ionized 
Fe K$_{\alpha}$ emission lines. \cite{2000Sordo} have investigated two BeppoSAX observations of Cen X-3 in August 
1996 and in February 1997.
They found significantly higher energy of the fluorescent
Fe K$_{\alpha}$ emission line ($\sim$6.6 keV), which could be a blend of 6.4 keV emission line with 6.7 keV 
emission line originating from hot matter surrounding the
neutron star. \cite{2012naik} with the same \textit{XMM-Newton} observation which we have analysed here,
found significant variability
in the line parameters of the three Fe K$_{\alpha}$ emission lines during eclipse and out-of-eclipse phases.
They also suggest that the colder material, causing 6.4 keV line emission, is relatively close to the 
neutron star at most on the scale of the companion, while the hot matter producing the Fe XXV and Fe XXVI emission lines 
are further away as suggested by \cite{1996Ebisawa}.
A cyclotron resonance scattering feature at 28 keV and 30 keV has been detected in Cen X-3
with two \textit{BeppoSAX} observations \citep{1998Santangelo, 2000Burderi}.
\textit{Suzaku} observation of Cen X-3 covering a full binary orbit showed multiple
extended dips with spectral characteristics similar to that of the source in eclipse, indicating 
the dips to be produced due to obscuration by dense matter which are structures in the outer
region of the accretion disk \citep{2011Naik}
\par
LMC X-4 is a wind and disk fed persistent system in the Large Magellanic Cloud (LMC) which often shows 
X-ray flares sometimes with super Eddington luminosity \citep{2000Levine541, 2003Moon}. 
The source was found to show periodicity longer than the orbital period known as superorbital period of 30.5 
days \citep{1981Lang}. 
The pulsation of the soft spectral component and the powerlaw component showed significant phase differences
with the \textit{ROSAT}, \textit{GINGA}, \textit{ASCA}, \textit{BeppoSAX} and \textit{XMM-Newton} and \textit{Suzaku}
observatories \citep{1996Woo, 2002Paul, 2004naikandPaul, 2017-LMC-Beri, 2010Hung}.
These indicate different origin of the soft and hard X-rays.
LMC X-4 shows short super-Eddington bursts, during which the pulse profile of the pulsar 
changes both in phase and shape. Same has been 
observed with RXTE-PCA \citep{2000Levine541},  \textit{XMM-Newton} \citep{2017-LMC-Beri} and  \textit{NuStar} \citep{2018Shtykovsky}. 
\par
SMC X-1 is a wind and disk fed binary pulsar located in the Small Magellanic Cloud (SMC),
which has been observed with both at sub-Eddington and super-Eddington luminosities
and shows short and intense bursts like LMC X-4 \citep{1971Price, 1973Ulmer, 1981Coe}.
It has a highly variable superorbital period (between $\sim$40-65 days, \citealt{2013Hu}). 
\textit{ROSAT}, \textit{GINGA} and \textit{ASCA}
observations of the source show  out-of-eclipse flux variation by a factor of 20 between high and low state while
eclipsed flux and spectral parameters remain nearly same in both the states.
This possibly indicates blocking of the  
neutron star's  direct emission by a precessing tilted accretion disk \citep{1998Wojdowski}.
Observations of SMC X-1 have been carried out at different superorbital phases with \textit{Chandra} 
and \textit{XMM-Newton} to investigate the case of a precessing warped accretion disk in this source \citep{2005Hickox}.
Boradband spectroscopy of SMC X-1 carried out with multiple \textit{Suzaku} \citep{2018Kubota} and NuStar \citep{2019Pike}
 observations have reinforced the same.
\par
4U 1700$-$377 is a high mass X-ray binary wind fed system in which the nature of the compact object is not yet clear. 
The source has been observed with pronounced variability with strong flares,
which is believed to be due to accretion from the supergiant's 
inhomogeneous wind \citep{1983White}.\cite{2005van_der_Meer}
 have detected several recombination and fluorescent lines
including highly ionized and fluorescent Fe K$_{\alpha}$ emission lines in the eclipse and eclipse 
egress and low flux \textit{XMM-Newton} spectra, where lines are most prominent during eclipse.
They suggest extended ionized plasma
surrounding the compact object.  \cite{2015Jaisawal}
 have found 6.4 keV Fe K$_{\alpha}$ and 7.1 keV Fe K$_{\beta}$ emission lines
in 20 time resolved out-of-eclipse \textit{Suzaku} spectra of the source. The relation between the 
flux and equivalent width of these Fe emission lines indicate that these lines originate
from the matter near the compact object. They found a significant increase of the line of sight 
equivalent Hydrogen column density during low flux out-of-eclipse phases, indicating an inhomogeneous
distribution of the wind.
\par
4U 1538$-$522 is a wind fed X-ray binary pulsar. 
\cite{2011Roca} have analyzed eclipse and phase averaged out-of-eclipse spectra 
with \textit{XMM-Newton}. They 
found Fe emission lines in the energy range of 6--7~keV and several recombination lines 
below 3~keV in the eclipse spectrum. These indicate presence of emitting
material over a large distance comparable or greater than the size of the supergiant companion. \cite{2006Uddi}
 have detected a 6.4 keV Fe emission line with \textit{RXTE} and \textit{BeppoSAX}.
They have found varying line of sight equivalent Hydrogen column density over the orbital phase which
increases before and after the eclipse asymmetrically. This indicates a trailing accretion stream
or wakes from the supergiant companion. A cyclotron resonance scattering feature at around
22 keV has been observed in 4U 1538$-$52 in detail with \textit{Suzaku} \citep{2014Hemphill}, \textit{AstroSat} \citep{2019Varun}
and \textit{NuStar} \citep{2019Hemphill}.
\par
IGR J18027$-$2016 is a wind fed high-mass X-ray binary pulsar. 
\textit{INTEGRAL} and \textit{XMM-Newton} observations of the source show a high
value of the line of sight Hydrogen column density, which indicates intrinsic absorption \citep{2005Hill}.
\cite{2016Aftab}  have found several 
low intensity states in the source with \textit{Swift}-XRT, while \textit{Swift}-BAT
shows smooth variation over the orbit.
They have also found high value of line of sight Hydrogen column density, which even increases more before and after
the eclipse. These indicate an accreting trailing wind from the supergiant companion to the pulsar 
crossing the line of sight like in 4U 1538$-$522.
A cyclotron resonance scattering feature at 23 keV
has been discovered in IGR J18027$-$2016 with \textit{NuStar} \citep{2017Lutovinov}.
\par
IGR J17252$-$3616 is a wind fed X-ray binary pulsar 
showing strong X-ray absorption 
\citep{2004Walter, 1989Tawara}.
\cite{2007Thompson} have obtained orbital phase resolved spectroscopy with \textit{RXTE}.
They have found variation in flux, line of sight equivalent Hydrogen column density,  photon index
and Fe line equivalent width. Column densities rise by a factor of $\geq$10 
just before the eclipse ingress and after the egress, this indicates spherically symmetric wind
outflow from the supergiant which is trailing the pulsar in it's orbit.
\par
IGR J16479$-$4514 and IGRJ 16418$-$4532 are SFXTs discovered with \textit{INTEGRAL} in 2003
\citep{2006Negueruela, 2005Sguera}.
 IGR J16479$-$4514 showed more frequent X-ray outburst than other SFXTs in the past \citep{2007Walter_4514}.
  It is the SFXT with smallest orbital period of 3.32 days \citep{2009Jain}.
 Accretion onto the compact object is very likely from the supergiant's wind \citep{2008Romano}.
  The X-ray emission from the source was observed to be highly variable on time-scales of seconds to weeks 
during outbursts and in quiescence \citep{2008Sidoli}. 
Occurence of flares in similar orbital phases, suggests phase locked wind structure in the system 
\citep{2009Bozzo, 2013Sidoli}.
IGR J16418$-$4532 is a heavily absorbed pulsar as observed with \textit{INTEGRAL} \citep{2006Walter}.
A superorbital modulation has been observed in the source \citep{2013Corbet_4532, 2013ATelDrave_4532}.
It is an intermediate accretor between pure wind
accretion and full Roche lobe overflow \citep{2012Sidoli_4532}.
\cite{2013drave} with an analysis of \textit{XMM-Newton} data suggests clumpy 
wind structures of the supergiant's wind. 
\par
We have also analyzed the out of eclipse spectra,
whenever available in the same observation along with the
eclipse. We try to infer the X-ray wind characteristics of these
systems, which is the main reprocessing agent in the high
mass X-ray binaries.

\section{Observation and data analysis}
\textit{XMM-Newton} was launched in December 1999. The X-ray observatory consists of 3 sets of co-aligned 
X-ray telescope each with an effective area of 1500 cm$^{2}$ with the following focal plane 
instruments: three European Photon Imaging Camera (EPIC) and two Reflection Grating Spectrometers (RGS).
Two of the cameras are made of Metal Oxide Semi-conductor (MOS) CCD \citep{2001turn}
arrays and one uses pn CCDs  \citep{2001struder}.
The pn camera is placed at the focal plane of one telescope and the two MOS cameras along with the RGS  are
placed at the  focal plane of the other two telescopes. There is one co aligned Optical/UV Monitor (OM) telescope   \citep{2001mason}
which provides simultaneous optical and UV coverage with the X-ray instruments.
\par
According to the observation requirements different read out modes can be selected \citep{2012lumb}, 
namely full frame; large window; small window; and
Timing mode. The MOS and pn cameras provide imaging 
over the 30$'$ field of view in the energy range of 0.15--15 keV with a spectral 
 resolution of 20--50 (E/$\Delta$E) and angular resolution of 6$''$. 
 \par
 The Burst Alert Telescope (BAT) \citep{barthelmy2005} onboard the \textit{Swift} mission 
 \citep{gehr2004} observatory is a coded mask  aperture instrument with CdZnTe (CZT) detector with a 
  field of view (FOV) of $100^\circ$$\times$$60^\circ$. It operates in the energy range of 
  15-150 keV with a detection sensitivity of 5.3 mCrab in one day of observation time \citep{krimm2013}.
  We have used long term satellite orbit-wise light curves to identify the eclipse phases of the
  eclipsing HMXBs.

\par
We used the catalogue of \cite{2000liu}  to select eclipsing HMXBs  
and then searched for the available EPIC pn observations.
We shortlisted all EPIC pn observations covering eclipse
comparing with the long term \textit{Swift}-BAT orbital
profiles (Left figures in Figure \ref{fig:cenx3}-\ref{fig:IGR3616_1001}). 
We found eclipse observations of 9 HMXBs with a total of 13 observations, 
11 of which were carried out in the imaging mode,
the other two are with Timing mode. The details of
the mode of operation are given in Table \ref{mode}.
We used the \textit{XMM-Newton} Science Analysis Software (SAS) version
14.0.0 to reduce the data. 
\begin{table*}
  \caption{The \textit{XMM-Newton} EPIC pn log of observations of the HMXBs}
 \centering
 \begin{tabular}{lc cc crr}
 \hline
 Source 	   & Observation ID & Date	 	 &   Datamode    & 	Submode 		& Effective & Average\\
		   &     	    & of	 	 &     	 	 &				& Exposure  & count-rate	\\ 
		   &     	    & observation	 &     	 	 &				& (s)	    &   (c/s) \\ 
                   &                &                    &               &                              &           &  (OOE) \\
 \hline
 Cen X-3          &  0111010101    & 27-01-2001	 &   IMAGING   	 &	PrimeSmallWindow	        & 67250	    &41.37\\
 LMC X-4          &  0142800101    & 09-09-2003    	 &   IMAGING   	 &  	PrimeSmallWindow	& 113171    &51.05 \\
 LMC X-4          &  0203500201    & 16-06-2004 	 &   TIMING  	 & 	FastTiming		& 41360	    &43.21\\
 SMC X-1          &  0011450101       & 31-05-2001 	 &   IMAGING 	 & 	PrimeFullWindow		& 56389     &54.57\\  
 4U 1700$-$377       &  0083280401    & 20-02-2001 	 &   TIMING   	 &  	FastBurst		& 30448     &22.05\\
 4U 1700$-$377       &  0600950101    & 01-09-2009 	 &   IMAGING	 & 	PrimeFullWindow		& 49533     &-\\
 4U 1538$-$522       &  0152780201    & 14-08-2003	 &   IMAGING  	 & 	PrimeFullWindow		& 79035     &10.33\\
 IGR J18027$-$2016   &  0745060401    & 11-09-2014	 &   IMAGING  	 & 	PrimeFullWindow		& 43041     &2.89\\
 IGR J17252$-$3616   &  0405640201    & 29-08-2006	 &   IMAGING 	 & 	PrimeLargeWindow	& 20672	    &-\\
 IGR J17252$-$3616   &  0405640601    & 08-09-2006	 &   IMAGING 	 & 	PrimeFullWindow		& 12104     &-\\
 IGR J17252$-$3616   &  0405641001    & 27-09-2006	 &   IMAGING  	 &	PrimeLargeWindow	& 10172     &- \\
 IGR J16479$-$4514   &  0512180101    & 21-03-2008 	 &   IMAGING  	 & 	PrimeSmallWindow	& 32752     &-	\\
 IGR J16418$-$4532   &  0679810101    & 01-09-2012	 &   IMAGING 	 & 	PrimeLargeWindow	& 18281     &-	\\
\hline 
 \end{tabular}
  \label{mode}
  \\
  
 \footnotesize{OOE: Out-of-eclipse}
 \end{table*}
%
We followed Guainazzi et al. 2014\footnote[1]{XMM-Newton Calibration Technical Note 
(XMM-SOC-CAL-TN-0083, Nov 14, 2014)} and \cite{2006Kirsch}
for the data reduction and region selection
 for the observations with TIMING
mode.
We followed {\small SAS} 
threads\footnote[2]{https://www.cosmos.esa.int/web/xmm-newton/sas-thread-timing}$^{,}$\footnote[3]{https://www.cosmos.esa.int/web/xmm-newton/sas-thread-pn-spectrum}$^{,}$\footnote[4]{https://www.cosmos.esa.int/web/xmm-newton/sas-thread-pn-spectrum-timing}
for the extraction of the lightcurve and spectra from the cleaned event of the data.
We  extracted the event files 
with the {\small SAS} tool {\small  EVSELECT}. We checked for flaring particle background and did not find it 
in any of the observations.
We extracted the events in the energy range 
of 0.3--12 keV.
We first extracted the lightcurve for the whole duration of each observation, 
then identified the eclipse and out-of-eclipse (whenever available) duration comparing it with the long term average BAT orbital profile
and extracted the eclipse and out-of-eclipse events.
From these events we extracted source along with background images for each 
observation. Seeing the image quality we decided the size of the 
source region to avoid contribution from excess background counts and edge of a CCD. 
In case of observations with imaging mode we extracted (18--30)$''$ circular source regions.
For each observation we extracted circular background 
region of same size of the source region from a region which is free from any other X-ray sources.
In case of Timing mode we  extracted box regions with 33$\leq$RAWX$\leq$42 
for the source and 3$\leq$RAWX$\leq$5 for the background for one observation (ID 0203500201) of LMCX-4,
 with 28$\leq$RAWX$\leq$42 for the source and 4$\leq$RAWX$\leq$6 for the background for one observation (ID 0083280401) of 4U 1700$-$377.
The Timing mode observation of 4U1700$-$377 was carried out with FastBurst submode. For this observation we extracted the regions
with RAWY$\leq$140 (\citealt{2006Kirsch}).
\par
Many times the data is affected by pile-up. Pile-up causes energy of two photons detected within
one CCD exposure to be added, thus it reduces the number of soft X-ray photons and increases the number of hard X-ray photons,
leading to an artificial hardening of the spectrum. Hence correction for the pile-up is essential. 
For EPIC pn observations with FullWindow, LargeWindow, SmallWindow modes the maximum count-rate
above which pile-up becomes important are 2(4), 3(6), 25(50) for 2.5(5)$\%$ flux loss \citep{2012Jethwa} respectively.
However, it is recommended to use the {\small  SAS} task {\small EPATPLOT}, to check the effect of pile-up in the  
data\footnote[5]{https://www.cosmos.esa.int/web/xmm-newton/sas-thread-epatplot}.
We did not notice pile-up in the eclipse data, as the count rate was quite low, compared to the out-of-eclipse phases.
For some of the observations during out-of-eclipse phases
we  noticed pile-up (Cen X-3, OB ID: 0111010101; LMC X-4, OB ID: 0142800101; SMC X-1, OB ID:  0011450101), with the application of the 
task {\small EPATPLOT}. 
The central part is most likely to be affected by pile-up for pointed observations.
For the observations affected by pile up, 
 we first removed the part of the point spread function affected by pile-up, by removing pixels of central 5$''$ region
 and obtained spectra. 
 We then extracted spectra from an annular source region of inner radius 10$''$. 
 We found stable spectral parameters for same best fit model in both the cases, then finally 
 we obtained the spectra with annular source regions of  5$''$ inner radius
 for the observations in which we  noticed pile-up.
\subsection{EVENT SELECTION FROM THE LIGHTCURVES}  
We extracted source and background lightcurves from the source and background region files respectively from 
the single and double events 
(with PATTERN$\leq$4) for the  full exposure time for each observation with the {\small  SAS} task {\small  EVSELECT}. We
 subtracted the background lightcurve from the source lightcurve to obtain background
subtracted source lightcurve using the {\small FTOOL} {\small LCMATH}.
We confirmed the eclipse and the out-of-eclipse phases (whenever available)
by plotting the EPIC pn lightcurve along with the average orbital profile of the 
long term \textit{Swift}-BAT lightcurve \citep{2013Krimm}
for each observation. In the left panels of Figure \ref{fig:cenx3}--\ref{fig:IGR3616_1001} the top panels 
show the long term \textit{Swift}-BAT  orbital profiles and the bottom panels the EPIC pn
lightcurves. We mark the duration of eclipse events within two solid lines and
that of the out-of-eclipse persistent phases within two dashed lines.

Then we extracted separately  1) eclipse events 2) out-of-eclipse persistent events from the 
event files, for the observations which cover both the eclipse and the out-of-eclipse phases.
The observations which were taken during eclipse phases only, we extracted eclipse spectra from the 
full event files of those observations. To generate the eclipse and out-of-eclipse event files 
we first noticed the start and end time of 
each phases, then used the {\small  SAS} task {\small  GTIBUILD} to generate good time intervals of the two phases.
Then using the  {\small  SAS} task {\small  EVSELECT} we extracted the eclipse and out-of-eclipse events (whenever available) 
for each observation. 
\subsection{SPECTRAL ANALYSIS}
We  extracted the spectra for both the eclipse and the out-of-eclipse phases (when available), 
from the single and double events 
(with PATTERN$\leq$4) excluding the events which are at the edge of a CCD
and at the edge to a bad pixel (with FLAG=0).
For the extraction of the spectra we  used the {\small  SAS}  task {\small  EVSELECT}. We  generated response and ancillary 
files using the {\small  SAS} task {\small  RMFGEN} and  {\small ARFGEN} respectively.
We then extracted spectra from these eclipse events for each observation
and from the events during out of eclipse persistent phases
whenever available. We  rebinned the spectra with 25 counts per bin with
the {\small SAS} routine  {\small SPECGROUP} and therefore used $\chi^{2}$ statistics. 
Observations with IDs 0405640201, 0405640601, 0405641001 of IGR J17252$-$3616 and 0679810101 of 
IGR J16418$-$4532 and 0600950101 of 4U 1700$-$377 were taken only during eclipse, 
observation 0512180101 of IGR J16479$-$4514 covers a small part 
of the ingress along with total eclipse phase. So for these 6 observations no out-of-eclipse spectra are available.  
We fitted the X-ray spectra using {\small XSPEC} v12.8.2.

\par
The out-of-eclipse spectra and the eclipse spectra are shown in the top panels of each figures at right in Figure 
(\ref{fig:cenx3}-\ref{fig:U1700_id401}) and in Figure (\ref{fig:U1538} and \ref{fig:IGR2016}) where the middle and
the bottom panels show the contribution  of each bin towards the $\chi$ for the out of eclipse and eclipse spectra 
respectively. The top right panel of Figure \ref{fig:spec_1700_ecp} and Figure (\ref{fig:IGR4514}-\ref{fig:IGR3616_1001}) 
show the eclipse spectra of these 6 observations where the bottom panels show the contribution of $\chi$ for each bin for
the eclipse spectra only.

\subsubsection{\textbf{ECLIPSE AND OUT-OF-ECLIPSE SPECTRA OF THE HMXBs}}
The eclipse spectra were primarily fitted with {\small XSPEC} model \textit{powerlaw} modified by photoelectric absorption
with model \textit{phabs} to account for the Galactic line of sight and local absorption
and a high energy cut-off ({\small XSPEC} model \textit{highecut}) for some sources. 
A few of the sources showed excess low energy 
emission and a blackbody component ({\small XSPEC} model \textit{bbodyrad}) was added to the model for these sources.
Excess narrow emission profiles were seen in some of the spectra at specific energies.
The excess at specific energies were fitted with Gaussian functions ({\small XSPEC} model gauss).
Some sources show a significant soft excess below 1.5 keV while the power-law component is highly absorbed. The soft excess, 
therefore cannot originate at the same location as the power-law for these observations. 
We have fitted the soft excess as a blackbody ({\small XSPEC} model \textit{bbodyrad}) with a 
different absorption column density in those observations (model \textit{phabs}), only for the purpose of characterizing 
its temperature and flux. 
If the soft excess originates in a much larger region elsewhere in the binary, its true nature may not be a blackbody
and a bremsstrahlung may be a more appropriate description. For the spectra in which soft excess could not be fitted
with blackbody, {\small XSPEC} model \textit{bremss} have been used.
We note that several eclipse spectra showed a very low even a slightly negative 
photon index in the 1-10 keV band. However, this  does not lead to large bolometric luminosity
as the accretion powered pulsars have an exponential cut-off, which is usually above
10 kev and outside the EPIC pn energy band.
 The details of the models can be found 
 online in HEASARC webpage\footnote[6]{https://heasarc.gsfc.nasa.gov/xanadu/xspec/manual/Models.html}.
The best fit parameters for the eclipse and the out-of-eclipse spectra for all eclipsing HMXBs and details of 
individual sources are discussed below.\\
\par
\textbf{Cen X-3:}\\
 \\
 \cite{1996Ebisawa} carried out spectral analysis of \textit{ASCA} data
 in pre-eclipse, ingress, eclipse and egress
 phases in two energy bands (0.7-4 and 5-10 keV). In the high energy band they obtained best fits with 
 a cut-off powerlaw modified with photoelectric absorption along the line of sight, 
 an Fe edge ($\sim$7.1-7.2 keV) and four Gaussian functions for Fe  K$_{\alpha}$ ($\sim$6.4 keV, 
 $\sim$6.7 keV, and $\sim$6.97 keV) and Fe  K$_{\beta}$ (fixed with 7.06 keV)  emission lines. 
 In low energy band, they fitted the above spectra with absorbed powerlaw with 4-6 emission 
 lines. The broad band (0.5-70 keV) eclipse and out-of-eclipse spectra of Cen X-3 with \textit{Suzaku}
 \citep{2011Naik} were well fitted with partial covering powerlaw model with high-energy
cutoff 
 and Gaussian functions for the three Fe emission lines (6.4 keV, 6.7 keV, and 6.97 keV). 
 \cite{2012naik} have obtained best fit with \textit{XMM-Newton} eclipse and out-of-eclipse spectra in 
 the energy range of (4-10 keV) with powerlaw and three Gaussian functions for the above
 three Fe emission lines. Powerlaw photon index were found to be the lowest in the eclipse phase 
 which then increases with the increase of flux.
\par
The EPIC pn observation of Cen X-3 (OB ID: 0111010101) covers a large fraction of the total eclipse, 
eclipse egress and a small part of the out-of-eclipse persistent phase.
The eclipse spectrum of this observation was best fitted  with  a powerlaw, a blackbody emission (bbodyrad), and 8 Gaussian functions for emission lines 
at 0.98, 1.43, 1.98, 2.62, 3.34, 6.41, 6.66, 6.94 keV modified by photoelectric 
absorption. The out-of-eclipse spectrum has been fitted with the same model used for
the best fit of the eclipse spectrum but with different model parameters. The details are given in Table 3, 4 and 5.
In the out-of-eclipse phase the powerlaw photon 
index is slightly smaller, total flux in the energy range of (0.3-10) keV  is larger by a factor of $\sim$10 compared
to the eclipse phase. We have detected 3 Fe emission lines (fluorescent Fe  K$_{\alpha}$, Fe XXV and Fe XXVI) 
in both the eclipse and the out-of-eclipse phases. The flux for the three lines are 0.88, 2.12, 1.65 respectively 
in the eclipse phase and 15.01, 11.64, 9.58 respectively in the out-of-eclipse phase in units of 
10$^{-4}$ photons cm$^{-2}$ sec$^{-1}$. The corresponding equivalent widths  are  $\sim$104$^{+9}_{-11}$  eV,
 $\sim$246$^{+10}_{-12}$ eV, $\sim$209$^{+11}_{-11}$ eV in the eclipse phase, and $\sim$189$^{+12}_{-8}$ eV, 
 $\sim$131$^{+5}_{-7}$ eV, 139$^{+7}_{-7}$ eV in the  
 out-of-eclipse phase respectively.
The Fe  K$_{\alpha}$ emission line flux increases more than 17 times and the corresponding 
equivalent width increases by a factor of $\sim$2 in the out-of-eclipse phase than the eclipse phase.
An increase of the equivalent width in the out-of-eclipse phase from the eclipse phase is not expected 
when the emitting material is distributed in an extended region.
The fluxes of the Fe XXV and Fe XXVI emission lines increase by a factor of more than 5 but their equivalent 
widths fall by a factor of  $\sim$2 in the out-of-eclipse phase compared to their values in the eclipse phase. 
An increase of equivalent width of fluorescent Fe K$_{\alpha}$ emission lines has been observed in the out-of-eclipse
phase compared to the eclipse phase earlier from the same \textit{XMM-Newton} observation \citep{2012naik}
and from an \textit{ASCA} observation of Cen X-3 \citep{1996Ebisawa}. \\

{\textbf{LMC X-4:}}\\
\\
\cite{2004naikandPaul} reported three different fits of the \textit{BeppoSAX} 
spectra in the low state of LMC X-4, amongst which the fit with powerlaw, 
a bremsstrahlung component (kT $\sim$0.4 keV) and a Gaussian function for 6.4 keV Fe
emission line, together modified with line of sight  photoelectric absorption 
is reported to be the best. The best fit for the high state of \textit{BeppoSAX}
spectra were obtained with two absorbed powerlaws (hard with $\Gamma$ $\sim$0.65, 
soft with $\Gamma$ $\sim$3), a blackbody component (kT $\sim$0.15 keV) and a Gaussian function
for 6.4 keV Fe emission line.
\cite{2009Neilsen} have obtained the \textit{XMM-Newton}-RGS and \textit{Chandra}-High-Energy 
Transmission Grating Spectrometer (HETGS)
spectra in hard and low state of LMC X-4 respectively. They obtained a best fit of the RGS spectra
with powerlaw, bremsstrahlung (kT $\sim$0.455 keV), blackbody (kT $\sim$0.043 keV) components 
and Gaussian functions for many emission lines of Nitrogen and Oxygen, all modified with
line of sight photoelectric absorption. 
HETGS spectra gave the best fit with a model comprises of powerlaw, 
bremsstrahlung (kT $\sim$0.43 keV) and Gaussian functions corresponding to different species
of emission lines of Neon, Nitrogen, Oxygen and three Fe K$_{\alpha}$ emission lines, modified with
photoelectric absorption
\par
We have analyzed two EPIC pn observations of LMC X-4, where first one was observed on 9th September 2003
and the second one was observed on
16th June 2004.
The first EPIC pn observation of LMC X-4 (OB ID: 0142800101) covers three pre-eclipse bursts, 
out-of-eclipse persistent emission, eclipse ingress,
full length of the total eclipse and eclipse egress. The other observation of this source
(OB ID: 0203500201) included out-of-eclipse persistent phase, eclipse ingress, full length
of the total eclipse including a small burst and eclipse egress phases. 
In the first EPIC pn observation (OB ID: 0142800101) of LMC X-4 there were three bursts at the beginning of the observation.
For the out-of-eclipse spectrum for this observation we excluded the period of the bursts and only used data
where the source intensity was steady, as shown by the dashed lines in the left panel of Figure \ref{fig:lmcx4_101}. 
The pulse profile evolution of LMC X-4 during and after the flares has been reported 
from this observation by \citet{2017-LMC-Beri}. In the other observation (OB ID: 0203500201) of LMC X-4 a small burst has been found
during eclipse, we  extracted the eclipse spectrum for this observation excluding the events during the bursts.
The eclipse spectrum of the first observation (OB ID: 0142800101) of LMC X-4 is fitted with a powerlaw,
a blackbody and a Gaussian function for the Fe  K$_{\alpha}$ emission line modified by line of sight
photoelectric absorption.
The X-ray spectrum of LMC X-4 shows very low absorption column density (N$_{\rm H}$). 
N$_{\rm H}$ could not be constrained with the \textit{XMM-Newton} spectrum.
The X-ray spectrum of LMC X-4
 detected down to 0.1 keV with the LECS instrument of \textit{BeppoSAX} \citep{2004naikandPaul}
with very low N$_{\rm H}$.
We have taken the Galactic column density of 0.06$\times 10^{22}$ cm$^{-2}$ towards LMC X-4 
(HEASARC TOOL\footnote[7]{https://heasarc.gsfc.nasa.gov/cgi-bin/Tools/w3nh/w3nh.pl})
 as lower limit of the absorption column density for the X-ray spectrum. 
The powerlaw photon index has been obtained to be 0.0$^{+0.3}_{-0.3}$ for the eclipse spectrum.
The best fit value for the energy of the fluorescent Fe K$_{\alpha}$
emission line is 6.36 keV with a flux of 0.07 $\times$10$^{-4}$ photons cm$^{-2}$ sec$^{-1}$.
 For the out-of-eclipse spectrum of this observation the best spectral fit is 
obtained with model consisting of a power-law, 
a blackbody emission, a bremsstrahlung
radiation and two Gaussian functions (one for Fe XXV emission line another 
for a low energy emission line), modified by photoelectric absorption.
N$_{\rm H}$ is frozen with the same value as for the eclipse spectrum.
\citet{2017-LMC-Beri} obtained similar parameters with this model, in the best fit of the phase averaged
persistent spectrum of the same EPIC pn observation.
In the out-of-eclipse phase, the power law photon index increases by  $\sim$0.7,
the total flux in the energy range of (0.3-10) keV is larger by factor of more than 237 compared to the eclipse phase.
  Lowly ionized (or neutral) Fe K$_{\alpha}$ emission line has been detected during eclipse with high equivalent width 
(597$^{+171}_{-256}$ eV) but it is absent in the out-of-eclipse phase (upper limit of equivalent width $\sim$20 eV), 
while highly ionized 6.60 keV  Fe  XXV emission line has been 
detected in the out-of-eclipse phase, with line flux of 4.03 $\times$10$^{-4}$ photons cm$^{-2}$ sec$^{-1}$ 
and equivalent width of 
$\sim$166$^{+12}_{-11}$ eV, but it is absent in the eclipse phase.
We have obtained a low energy emission line with energy 0.96$^{+0.01}_{-0.01}$ keV.
\cite{2009Neilsen} have detected
13.560 A$^\circ$ (0.91 keV) Ne IX He$_{\alpha}$ and 12.158 A$^\circ$ (1.02 keV) Ne X Ly$_{\alpha}$
emission line near 1 keV with the \textit{Chandra} HETGS.
The 0.96 keV line detected by EPIC pn could be a mix of the two lines at 0.91 keV and 1.02 keV
of Ne seen earlier with the HETGS instrument.
\par
The second observation of LMC X-4 (OB ID: 0203500201) was carried out in Timing mode.
The best fit model for the eclipse spectrum of this observation
is obtained with a powerlaw and a blackbody emission modified by photoelectric absorption.
The line of sight Hydrogen column density is fixed with 0.06$\times$10$^{22}$ cm$^{-2}$ for the eclipse spectrum similar to the
eclipse and the out-of-eclipse spectrum of the previous observation of this source.
The out-of-eclipse spectrum of this observation was best fitted with a powerlaw, 
a bremsstrahlung radiation and 3 Gaussian functions of
emission energies  0.98, 6.38, 6.99 keV, modified by photoelectric 
absorption with N$_{\rm H}$ frozen with a value of 0.06$\times$10$^{22}$ cm$^{-2}$. 
The powerlaw photon index in the out-of-eclipse phase 
increases by $\sim$0.8, while 
total flux in the energy range of (0.3-10) keV  is larger by a factor of 86 compared to the eclipse phase.
We have detected fluorescent Fe K$_{\alpha}$ emission line in the out-of-eclipse phase  with a flux of 
1.55$\times$10$^{-4}$ photons cm$^{-2}$ sec$^{-1}$ and the equivalent width of $\sim$117$^{+2}_{-2}$ eV. 
 Fe XXVI emission line has been detected 
in the out-of-eclipse phase  with a flux  of 
2.39$\times$10$^{-4}$ photons cm$^{-2}$ sec$^{-1}$ and an equivalent width of $\sim$195$^{+4}_{-3}$ eV
respectively. In the eclipse spectrum we found some positive residuals in (6-7) 
keV energy range, but due to poor 
statistics we could not constrain the energy of the emission lines. 
However  just to compare the line fluxes,
we fitted the eclipse spectrum with the two Fe emission lines (fluorescent Fe K$_{\alpha}$ and Fe XXVI),
freezing the line energy and width as obtained in the out-of-eclipse spectrum. 
Including these lines we have obtained very little improvement of 
$\chi$$^{2}$ from a value of 103.01 for 94 degrees of freedom to 99.93 with 92 degrees of freedom.
We have found a flux of 2.59$\times$10$^{-6}$ photons cm$^{-2}$ sec$^{-1}$ for fluorescent Fe K$_{\alpha}$
emission line and a flux of 2.80$\times$10$^{-6}$ photons cm$^{-2}$ sec$^{-1}$ for Fe XXVI  
emission line respectively, which
are less significant than 3$\sigma$. 
Like the out-of-eclipse phase of the previous observation, 
 0.98$^{+0.03}_{-0.03}$ keV emission line could be a sum of the two Ne lines
as detected by \cite{2009Neilsen} with \textit{Chandra} HETGS.
\\
 \par
{\textbf{SMC X-1:}\\
\\
\cite{2004Naik} have fitted the pulse phase average spectra with \textit{BeppoSAX} by
a hard powerlaw component with high energy cut-off, a blackbody emission for the soft excess and  
a Gaussian function for 6.4 keV Fe emission line modified by photoelectric absorption.
\cite{2013Hu} analyzed 
both the hard and soft state spectra of SMC X-1 with \textit{RXTE}-PCA. They best describe 
 the continuum of both the spectra with a Comptonized component,
 a blackbody emission and a Gaussian line of 6.4 keV Fe emission line. They found lower plasma 
temperature, Fe line width and equivalent width in the low state than in the hard state.
\par
The EPIC pn observation of SMC X-1 (Observation ID: 0011450101) covers a very small portion of the out-of-eclipse 
persistent phase, eclipse ingress and full length of the total eclipse.
The eclipse spectrum of  SMC X-1 has been modeled with a powerlaw with high energy cut-off, 
a blackbody emission and a Gaussian function for Fe
 K$_{\alpha}$ emission line of 6.38 keV modified by photoelectric absorption. The high energy cut-off was found 
 to have a value of 2.46 keV with a folding 
 energy of  3.6 keV. The  best fit for the out-of-eclipse spectrum is found with the above model except the 
  Fe  K$_{\alpha}$ emission line. The high energy cut-off is found to have a value of 2.59 keV with a folding 
 energy of 8.12 keV.  
 The powerlaw photon index has a negative value ($\Gamma$ = -0.56) in the eclipse phase and it increases to 0.2 in the
 out-of-eclipse phase. We also obtain a good fit for the eclipse spectrum with the same model as mentioned above for this phase
  with the same powerlaw photon index (0.2) as obtained for the best fit of the out-of-eclipse spectrum, but with the parameters
  for the  high energy cut-off and folding energy nearly doubling their values. 
$\chi^{2}$ (194.10) for this fit is higher than $\chi^{2}$ (181.73) of the previous fit.
 So we report the parameter values in Table \ref{NH} and \ref{iron} from the previous fit only.  
 The total flux in the energy range of (0.3-10) keV in the out-of-eclipse phase
 is larger by a factor of $\sim$77 compared to its value in the eclipse phase.
 During the eclipse phase we have detected a fluorescent
 Fe  K$_{\alpha}$ emission line with flux and equivalent width of
 0.07$\times$10$^{-4}$ photons cm$^{-2}$ sec$^{-1}$ and 126$^{+18}_{-18}$ eV respectively. We could not detect any 
 Fe  K$_{\alpha}$ emission line during the out-of-eclipse phase, but  we can not rule out the line as the
 duration of the out-of-eclipse phase compared to the eclipse phase is very short (lower by a factor of $\sim$44)
 and the continuum flux is higher by a large factor  
 ($\sim$77 times) in the out-of-eclipse  phase from the eclipse phase. 
 We derived an upper limit of equivalent width of $\sim$90 eV for an emission line of 6.4 keV. \\
 
 \par
{\textbf{4U 1700$-$377:}\\
\\
\cite{2005van_der_Meer} have obtained three different fits for the continuum of eclipse, eclipse egress and low flux spectra of 
4U 1700$-$377 with \textit{XMM-Newton}. The first model comprises of three powerlaw components, one for the direct 
X-rays from the compact object, another for the scattered X-rays from the surrounding medium and the third one is for
the soft excess. While the second model is described with first two powerlaws modified with high energy cut-off and in the 
third model powerlaw for the soft excess was replaced with a blackbody component. Several Gaussian functions were included for
the emission lines in all the spectra. The three fits gave similar reduced $\chi^{2}$. Amongst possible Fe emission lines 6.4 keV and 6.53 keV
Fe K$_{\alpha}$ and 7.06 keV Fe K$_{\beta}$ emission line have been found during eclipse, while 6.42 keV, 6.72 keV and 7.11 keV Fe
emission lines were observed in the egress and 6.41 keV, 6.72 keV (fixed) and 7.06 keV (fixed) Fe emission lines were detected 
in the low flux state. 6.53 keV Fe emission line seen during eclipse was suggested to come from higher ionized Fe ions (Fe XVIII-XXIV).
The relation between the ionization parameters for different lines and the line intensities indicate photoionized plasma surrounding
the central source. The fit of the soft excess could not give satisfactory explanation of its origin. Previously with \textit{GINGA},
the soft excess was fitted with bremsstrahlung component \citep{1992Haberl, 1994Haberl}, 
the temperature of which could not be constrained.
\par
We have analyzed two  EPIC pn observations of 4U 1700$-$377 (OB IDs: 0083280401, 0600950101) carried out on
20-21 February 2001 for $\sim$30 ks and on 1st September 2009 for $\sim$50  ks respectively.
The first one covers some part of the eclipse, the eclipse egress and a small portion 
of the out-of-eclipse persistent phase
and the second one covers nearly full length of the eclipse phase.
The first observation (OB ID: 0083280401) was carried out in Timing FastBurst mode.
   For the eclipse spectrum in this observation, the best fit has been
obtained with a
powerlaw and a Gaussian function with
energy 6.37 keV modified by photoelectric absorption. 
This observation includes a shorter duration of the eclipse and the spectrum
has limited statistics. 
The best fit for the out-of-eclipse spectrum of this observation is also obtained with
 the above model.
The powerlaw photon index has a  negative value (-0.41) during eclipse. 
In the  out-of-eclipse phase it increases by $\sim$
0.24 and is consistent with a range of (-0.72, 0.35).
  The total flux in the energy range of (0.3-10) keV is large by  $\geq$8
times compared to the eclipse phase.
During eclipse 6.37 keV Fe K$_\alpha$ emission line flux has been found to be
5.96$\times$10$^{-4}$ photons cm$^{-2}$ sec$^{-1}$ with a very high equivalent width of $\sim$1258$^{+101}_{-103}$ eV.
During out-of-eclipse phase the flux for 6.44 keV Fe K$_\alpha$ emission line
is found to be 14.88$\times$10$^{-4}$ photons cm$^{-2}$ sec$^{-1}$
which is $\sim$2.5 times higher than its value during eclipse and
the equivalent width decreases by a factor of $\sim$6.6 to $\sim$191 eV.
\par
The eclipse spectrum from the second pn observation (OB ID: 0600950101) has been fitted with a powerlaw
modified with high energy cut-off, blackbody emission
and 13 Gaussian functions with energies 0.81, 1.28, 1.75, 1.97, 2.34, 2.57, 2.99, 3.70, 4.12, 6.39,
6.68, 7.05 and 7.49 keV  modified by photoelectric absorption. 
The best fit is obtained with the addition of an edge at 7.06 keV.  The powerlaw photon index 
is found to have a negative value of -1.34.
 Value of the high energy cut-off is determined to be 5.58 keV with a folding
 energy of 4.59 keV.   
   The fluorescent Fe  K$_{\alpha}$ emission line flux has been found
   to be 4.89$\times$10$^{-4}$ photons cm$^{-2}$ sec$^{-1}$
   with very high equivalent width of $\sim$1061$^{+13}_{-13}$ eV. The Fe XXV emission line flux has been found to be low,
   0.26$\times$10$^{-4}$ photons cm$^{-2}$ sec$^{-1}$ with very small equivalent width of $\sim$22$^{+4}_{-3}$ eV.   
   7.05 keV  emission line flux and equivalent width have been found to be
   0.88$\times$10$^{-4}$ photons cm$^{-2}$ sec$^{-1}$ and $\sim$219$^{+25}_{-32}$ eV respectively.
\par
In the eclipse spectrum of the second pn observation we have detected many emission lines while in
the eclipse spectrum of the first observation we have detected only one emission line. The
eclipse duration covered in the first observation was significantly shorter than that in the second
observation. We checked for the upper limits of the emission lines which were not detected in the eclipse spectrum
of the first observation and we have found low upper limit of  equivalent width of those lines.
We can say, that because of limited statistics
we could not constrain  other emission lines in the eclipse phase of the first observation.
 The emission lines in 4U 1700$-$377 have been discussed by \cite{2005van_der_Meer}.
 In this work we restrict our discussion to the Fe lines that are common in most HMXBs.
\\

{\textbf{4U 1538$-$522:}\\
\\
\cite{2006Uddi} have obtined out-of-eclipse spectral analysis of the source with
\textit{RXTE} and \textit{BeppoSAX} at two different orbital phases.
They obtained best fit of \textit{RXTE} (3-10 keV) data with powerlaw
 with high energy cut-off and 6.4 keV Fe K$_{\alpha}$
emission line modified by photoelectric absorption.
While with \textit{BeppoSAX} in the energy range of 0.3-10 keV 
they obtained the best fit with the same model except the high energy cut-off.  
They have not seen any significant variation of the photon index and Fe line flux over the orbital
phase with orbital phase resolved spectral analysis while they have found large variation of 
column densities along the orbital phase which increases even more just before and after the eclipse asymmetrically.
They suggest nearly constant accretion rate and slightly assymetric condition of the accretion column 
due to trailing accreting material from the supergiant. 
\cite{2011Roca} have obtained orbital phase averaged and also the eclipse spectra with the same \textit{XMM-Newton}
observation of  4U 1538$-$522. They reported the best fit of the phase averaged spectra with three absorbed powerlaws
(with same photon indeices but different normalization and column densities) and 4 Gaussian functions for fluorescent $\sim$6.4 
keV Fe K$_{\alpha}$, $\sim$2.42 keV, $\sim$1.90 keV, $\sim$1.34 keV emission lines. They obtained best fit for the eclipse spectra
with two absorbed powerlaws and 6 Gaussian functions for  $\sim$6.4 keV fluorescent Fe K$_{\alpha}$, $\sim$6.63 keV Fe XXV, $\sim$2.44 keV,
$\sim$2.00 keV, $\sim$1.85 keV, $\sim$1.34 keV emission lines. They have identified the low energy emission lines 
from different species of Sulphur (S), Silicon (Si) and Mg (Magnesium) ions (Table 3 and 6, \citealt{2011Roca}). 
A soft excess was observed  below
0.5 keV, which could neither be satisfactory explained by a soft powerlaw nor by blackbody emission.
\par
The EPIC pn observation of 4U 1538$-$522 (OB ID: 0152780201) covers a small part of the
out-of-eclipse persistent phase, eclipse ingress
and full length of the total eclipse. 
The eclipse spectrum of this observation is best fitted with a 
powerlaw, 7 Gaussian functions with
energies 1.26, 1.85, 2.40, 6.02, 6.38, 6.69 and 6.95 keV modified by photoelectric absorption. 
The out-of-eclipse spectrum of this observation got the best fit with a
powerlaw and a Gaussian function for fluorescent Fe K$_{\alpha}$ emission line
together modified 
by photoelectric absorption along with a less absorbed 
blackbody emission. The line of sight equivalent Hydrogen column density
associated with the powerlaw emission is 17.16$\times$10$^{22}$ cm$^{-2}$,
while that for the blackbody emission is found to be 0.54$\times$10$^{22}$ cm$^{-2}$. 
This value is less by a factor more than 30 than that for the powerlaw emission.
The powerlaw photon index in the  out-of-eclipse phase increases by $\sim$0.5
and the total flux in the energy range 
of (0.3-10) keV is enhanced by a factor of $\geq$30
compared to its value in the eclipse phase.
Both the eclipse and the  out-of-eclipse spectrum show the signatures of  Fe  K$_{\alpha}$ emission line with 
fluxes of 0.57$\times$10$^{-4}$ photons cm$^{-2}$ sec$^{-1}$ and 2.46$\times$10$^{-4}$ photons cm$^{-2}$ sec$^{-1}$
with equivalent width of 792$^{+42}_{-42}$ eV and 79$^{+22}_{-18}$ eV respectively. The flux for the Fe  K$_{\alpha}$ emission line
increases by a factor of $\geq$4 while the equivalent width decreases by a factor of more than 10 in the out-of-eclipse 
phase from its value in the eclipse phase. Figure \ref{fig:U1538} shows that
the soft X-ray emission in  the eclipse and the out-of-eclipse spectra overlaps, corresponding flux in the energy
 range of 0.3-2.0 keV is obtained as 1.76$\times$10$^{-13}$ ergs cm$^{-2}$ sec$^{-1}$ and 2.08$\times$10$^{-13}$ 
 ergs cm$^{-2}$ sec$^{-1}$
respectively. The emission lines in 4U 1538$-$522 has been reported by \cite{2011Roca}.
 In this work we confine our discussion to the Fe lines only.\\

{\textbf{IGR J18027$-$2016:}\\
\\
\cite{2005Hill} carried out spectral analysis of IGR J18027$-$2016 with \textit{INTEGRAL} and \textit{XMM-Newton}.
With simultaneous fitting of both the spectra they obtained good fitting with an absorbed broken powerlaw
and a broad Gaussian for the soft excess. The break was found at $\sim$11 keV and the powerlaw photon
indices have values of $\sim$0.8 and $\sim$3. Addition of 6.4 keV and 7.1 keV emission lines slightly improves the fit.
With pulse phase resolved spectroscopy
they have found similar spectral parameters, with high value of 
line of sight Hydrogen column densities. These indicate that the 
absorption is not due to neutron star's accretion column and
the pulsar is surrounded by supergiant's dense wind.
\cite{2016Aftab} obtained orbital phase resolved spectroscopy of 33 \textit{Swift}--XRT observations and
found a large variation in the source flux, photon index and equivalent Hydrogen column density. This indicates
a strongly variable accretion rate of the pulsar and variable wind structures of the supergiant. The detection of several low 
intensity states along with high states with \textit{Swift}--XRT suggests clumpy wind structures or
hydrodynamic instabilities in the accreted material \citep{2016Aftab}.
\par
The EPIC pn observation of IGR J18027$-$2016  (OB ID: 0745060401) covers a major part of the total eclipse, eclipse egress and some 
part of the out-of-eclipse phase. 
The eclipse spectrum of this observation is best fitted with a powerlaw
 and two Gaussian functions, one for the fluorescent Fe  K$_{\alpha}$ emission line 
of energy 6.41 keV and another of highly ionized 6.66 keV Fe XXV emission line,
modified by photoelectric absorption. The N$_{\rm H}$ could not be constrained for the
eclipse spectrum. The flux for the Fe  K$_{\alpha}$ emission line is 
found to be 0.07$\times$10$^{-4}$ photons cm$^{-2}$ sec$^{-1}$ with an equivalent width of 445$^{+254}_{-191}$ keV.
The best fit for the  out-of-eclipse spectrum has been obtained with a powerlaw
modified by photoelectric absorption only.
The powerlaw photon index in the eclipse phase has been found to be
-0.22$^{+0.24}_{-0.23}$, which is consistent with a range of (-0.02, 0.54).
In the out-of-eclipse phase it increases by $\sim$1 and becomes positive (0.71).
The total flux in the out-of-eclipse phase in the energy range of (0.3-10) keV is
increased by a factor of $\geq$43. \\
\\
\textbf{IGR J17252$-$3616:}\\
\\
\cite{2011Manousakis} have fitted the orbital phase resolved \textit{XMM-Newton} spectra with intrinsically absorbed cut-off
powelaw, a blackbody radiation and a Gaussian function for 6.4 keV emission line. The cut-off energy was found to be
8.2 keV with photon index 0.02. The  blackbody temperature was 0.5 keV. They have also found Fe K-edge at 7.2 keV.
They obtained high and variable intrinsic absorbing column density (9-89$\times $10$^{22}$ cm$^{-2}$). 
Fe emission line equivalent width was found to be highly variable, in the range of 40-1100 eV.
\cite{1989Tawara} have found strong and variable absorption of the source fitting \textit{GINGA} data
with absorbed powerlaw. They found column density upto $\sim$10$^{24}$ cm$^{-2}$. They suggest 
dense matter surrounding the pulsar. With \textit{GINGA} data, taken $\sim$6 months later, \cite{1990Takeuchi}
found a lower value of the column density ($\sim$10$^{23}$ cm$^{-2}$), fitting the spectrum with a power law
modified with high energy cut-off. Pulse phase resolved spectroscopy of the source with \textit{INTEGRAL}
obtained by \cite{2006Heras} shows nearly the same continuum emission. 
\cite{2007Thompson} have fitted the several orbital phase resolved \textit{RXTE}
spectra including eclipse
with absorbed powerlaw modified by high energy
cut-off and a Gaussian function for 6.4 keV Fe emission line. They obtained cut-off of $\sim$16 keV 
for the best fit. They obtained column densities in the range of 8-126$\times $10$^{22}$ cm$^{-2}$,
photon indices in the range of 0.5-2 and huge variation in the 6.4 keV Fe emission line equivalent
width (133-6490 eV).
\par
We have analyzed three  EPIC pn observations of IGR J17252$-$3616, all during  eclipse phases observed 
between 29th august-27th September, 2006 (OB IDs: 0405640201, 0405640601, 0405641001).
All three observations were initially fitted with powerlaw modified by photoelectric absorption.
For the best fit, the first observation needed two Gaussian functions and
 a less absorbed blackbody emission 
One Gaussian is for Fe K$_{\alpha}$ emission line of 6.4 keV and another with a value of 7.01 keV.
The second observation was best fitted with the same model as the first observation, except the Gaussian
with a value of 7.01 keV. The third observation obtained the best fit with a Gaussian function for
Fe K$_{\alpha}$ emission line of energy 6.4 keV. The soft excess in this 
observation could not be fitted with blackbody emission. 
The power-law photon index is found to have values of 0.24$^{+0.83}_{-0.75}$, 0.81$^{+4.92}_{-1.17}$
and -0.52$^{+4.35}_{-2.47}$ respectively in the 
three observations, while the Fe K$_{\alpha}$ emission line flux is 0.13, 0.18, and 0.04 respectively in units of 
10$^{-4}$ photons cm$^{-2}$ sec$^{-1}$.
The equivalent width is found to be significantly different for the three observations with values 
of 2695$^{+415}_{-622}$ eV, 1831$^{+407}_{-305}$ eV and 921$^{+691}_{-460}$ eV respectively. 
The flux and equivalent width associated with 7.01 keV emission line is
0.05$\times$10$^{-4}$ photons cm$^{-2}$ sec$^{-1}$ and 1227$^{+982}_{-245}$ eV respectively in the first observation.
This is significant $\gtrapprox$ 3 $\sigma$.
We could not detect this line in the other two observations, as the data is not that good.
The line of sight equivalent Hydrogen column density for the power law component for the three observations
are 9.34, 18.17 and 13.48 respectively while that for the blackbody component are 0.90 and 2.67
 for the first two observations respectively (all values in units of $10^{22}~ \rm cm^{-2}$) 
 which are lower compared to
the same for the powerlaw component by a factor of $\textgreater$10 and $\sim$7 respectively. 
\\

{\textbf{IGR J16479$-$4514:}\\
\\
\cite{2009Romano} have fitted the out of out-burst \textit{Swift}-XRT data with an absorbed powerlaw plus blackbody emission.
They suggest a small region on the neutron star surface, perhaps polar cap as the origin of the blackbody emission.}
\cite{2013Sidoli} have observed the source with \textit{Suzaku} in eclipse and out-of-eclipse phase. They fitted 
all the spectra with absorbed powerlaw and 6.4 keV Fe emission line. They found hard powerlaw when the source is 
brighter like other SFXTs and HMXB pulsars. The line of sight coulumn density
did not show much variation outside eclipse.
The Fe line flux was observed to vary through out the orbit with a correlation with the unabsorbed
source flux 
above 7 keV. Fe line equivalent width was found to increase during eclipse as expected.
\par
The EPIC pn observation of IGR J16479$-$4514 (OB ID: 0512180101) covers the eclipse ingress and a major 
part of the total eclipse. We extracted only eclipse spectrum from the total
eclipse event for this observation.
The eclipse spectrum of this observation is best fitted with a powerlaw,
two Gaussian functions with energies 6.41 keV and 6.88 keV  
 modified by photoelectric absorption and a less absorbed blackbody emission
 associated with the line of sight equivalent Hydrogen column density
of 0.58$\times$10$^{22}$ cm$^{-2}$ which is a factor of $\sim$8 lower than that associated with the powerlaw. 
The powerlaw photon index is obtained to be 1.28$^{+0.37}_{-0.34}$.
The Fe K$_{\alpha}$ emission line flux is low with a value of 0.09$\times$10$^{-4}$ photons cm$^{-2}$ sec$^{-1}$ 
and  equivalent width of 803$^{+89}_{-89}$ eV. The energy value for the 6.88 keV  emission line is not well
constrained like Fe XXVI emission line for other observations (Please see Table \ref{iron}), the flux for this emission
line is 0.05$\times$10$^{-4}$ photons cm$^{-2}$ sec$^{-1}$ with an equivalent width of 576$^{+345}_{-230}$ eV.\\

{\textbf{IGR J16418$-$4532:}\\
\\
\cite{2012Romano} obtained the best fit of  \textit{Swift}-XRT out-of-eclipse 
spectra with single absorbed powerlaw.
They obtained high intrinsic absorption (upto $\sim$7$\times$10$^{22}$ cm$^{-2}$).
The spectral parameters did not show much variation though the flux variation was quite large. 
 Average spectra based on 2004 and 2011 \textit{XMM-Newton} data were well fitted with an absorbed powerlaw
 model \citep{2012Sidoli_4532}. The 2004 data showed marginal evidence of a soft excess below 2 keV.  
 They did not find clear evidence of any Fe emission line. 
  The eclipse spectrum with \textit{INTEGRAL} was fitted with a simple power law with photon idex of 2.2 \citep{2013drave}.
 The best fit of the mid-eclipse spectrum with \textit{XMM-Newton} was obtained with an absorbed powerlaw and
 two Gaussian functions for 6.4 and 6.65 keV Fe emission lines. The post eclipse spectra did not show any evidence of 
 an Fe emission line \citep{2013drave}.
 \par
The EPIC pn observation of IGR J16418$-$4532 (OB ID: 0679810101) covers only a major part of the total eclipse.
So for this observation we extracted the eclipse spectrum only, which has been fitted with a powerlaw
 and two Gaussian functions, one for fluorescent Fe K$_{\alpha}$ emission line of 6.43 keV and another 
 for highly ionized Fe XXVI emission line of 6.94 keV,  
modified by photoelectric absorption. 
The powerlaw photon index has been found to be 0.40$^{+0.83}_{-0.45}$.
The Fe K$_{\alpha}$ emission line flux is 0.06$\times$10$^{-4}$ photons cm$^{-2}$ sec$^{-1}$ with a high
equivalent width of 1358$^{+453}_{-453}$ eV. The Fe XXVI emission line flux has been found to be 
0.05$\times$10$^{-4}$ photons cm$^{-2}$ sec$^{-1}$ with equivalent width of 1911$^{+1146}_{-764}$ eV.\\
\\
{\textbf{Emission lines other than Fe K$_{\alpha}$:}\\
\\
We have observed emission lines other than Fe K$_{\alpha}$ in Cen X-3, LMC X-4, 4U 1700--377, 4U 1538--522 and IGR J17252--3616.
The line parameters and their previous detections have been given in Table \ref{elements_alltogether}.
Some of the emission lines were not observed earlier. However, in this work we focus our discussion on Fe emission lines only,
so details about other emission lines are not discussed here.
%
%
\begin{table*}[ht]
  \caption{The exposure times for the eclipse and the out-of-eclipse spectra of the HMXBs and the best fit spectral models}
  \centering
 \begin{tabular}{lccrcl}
 \hline 
 Source 	   & Observation ID 	  &  Phase  & exposure  & & Best fit model 		\\
  		   &                      &         &   (s)	 &&    \\
 \hline
 Cen X-3           &  0111010101          &  E      & 7152       &    & phabs$_{\small 1}$$\times$(bbodyrad$_{\small 2}$ + gaussian$_{\small 3}$   \\
 		   &                      &         &        	 && + gaussian$_{\small 4}$ + gaussian$_{\small 5}$ + gaussian$_{\small 6}$\\
 		   &                      &         &          &	& + gaussian$_{\small 7}$ + gaussian$_{\small 8}$  + gaussian$_{\small 9}$\\
 		   &                      &         &          &	&   + gaussian$_{\small 10}$ + powerlaw$_{\small 11}$)\\
  		   &                      &         &         	 &&  \\
 		   
 Cen X-3          &  0111010101          &  OOE    &  8937     &	&  phabs$_{\small 1}$$\times$(gaussian$_{\small 2}$ + gaussian$_{\small 3}$  \\
 		   &                      &         &       	  &&  + bbodyrad$_{\small 4}$  + gaussian$_{\small 5}$ + gaussian$_{\small 6}$ \\ 
 		   &                      &         &          &	&  + gaussian$_{\small 7}$ + gaussian$_{\small 8}$  + gaussian$_{\small 9}$\\ 
  		   &                      &         &   	  &&    + gaussian$_{\small 10}$ + powerlaw$_{\small 11}$)\\
 \hline
 LMC X-4          &  0142800101          &  E      &  8972   	  && phabs$_{\small 1}$$\times$(powerlaw$_{\small 2}$ + bbodyrad$_{\small 3}$  + gaussian$_{\small 4}$)  \\
  		   &                      &         &      	  &&    \\
		   
 LMC X-4          &  0142800101          &  OOE    &  38170     & 	& phabs$_{\small 1}$$\times$(powerlaw$_{\small 2}$ + bremss$_{\small 3}$ \\ 
 		   &                      &         &           & 	&  + bbodyrad$_{\small 4}$ + gaussian$_{\small 5}$ + gaussian$_{\small 6}$) \\
		   
\hline
 LMC X-4          &  0203500201          &  E      &  11630  	  &&   phabs$_{\small 1}$$\times$(bbodyrad$_{\small 2}$ + powerlaw$_{\small 3}$)\\
		   &                      &         &      	  &&      \\
 LMC X-4          &  0203500201          &  OOE    &  6336   	  &&  phabs$_{\small 1}$$\times$(bremss$_{\small 2}$ + gaussian$_{\small 3}$\\
		   &                      &         &    	  &&  + gaussian$_{\small 4}$ + gaussian$_{\small 5}$ + powerlaw$_{\small 6}$) \\
 \hline

 SMC X-1         &  0011450101          &   E     &  30790   	  &&  phabs$_{\small 1}$$\times$(highecut$_{\small 2}$$\times$powerlaw$_{\small 3}$\\
                   &                      &         &      	  && + bbodyrad$_{\small 4}$ + gaussian$_{\small 5}$)\\
                   &                      &         &      	  &&      \\    
 SMC X-1         &  0011450101          &   OOE   &  808 	  &&  phabs$_{\small 1}$$\times$(highecut$_{\small 2}$$\times$powerlaw$_{\small 3}$  + bbodyrad$_{\small 4}$)\\ 
 \hline

 4U 1700$-$377       &  0083280401          &   E     &  	622  &	&   phabs$_{\small 1}$$\times$(powerlaw$_{\small 2}$ + gaussian$_{\small 3}$) \\
                   &                      &         &           &   &     \\
 4U 1700$-$377       &  0083280401          &   OOE   &   34 	&  &   phabs$_{\small 1}$$\times$(powerlaw$_{\small 2}$ + gaussian$_{\small 3}$) \\
 \hline

 4U 1700$-$377       &  0600950101          &   E     &   42790    &	&  edge$_{\small 1}$$\times$phabs$_{\small 2}$$\times$(bbodyrad$_{\small 3}$ \\
                   &                      &         &     	&   & + gaussian$_{\small 4}$+ gaussian$_{\small 5}$ + gaussian$_{\small 6}$ \\
                   &                      &         &     	&   & + gaussian$_{\small 7}$ + gaussian$_{\small 8}$ + gaussian$_{\small 9}$ \\ 
                   &                      &         &      	&    &+ gaussian$_{\small 10}$ + gaussian$_{\small 11}$ + gaussian$_{\small 12}$\\
                   &                      &         &      	&    & + gaussian$_{\small 13}$ + gaussian$_{\small 14}$ + gaussian$_{\small 15}$ \\
                   &                      &         &      	&    &+ gaussian$_{\small 16}$ + highecut$_{\small 17}$$\times$powerlaw$_{\small 18}$)\\
 \hline

  4U 1538$-$522      &  0152780201          &   E     &    44780   &	&  phabs$_{\small 1}$$\times$(powerlaw$_{\small 2}$ + gaussian$_{\small 3}$ \\
                   &                      &         &      	&   &  + gaussian$_{\small 4}$ + gaussian$_{\small 5}$ + gaussian$_{\small 6}$ \\
                    &                     &         &        	&   & + gaussian$_{\small 7}$ + gaussian$_{\small 8}$ + gaussian$_{\small 9}$) \\
                   &                      &         &      	&   &\\                 
 4U 1538$-$522       &  0152780201          &   OOE   &   3171  	  &&  phabs$_{\small 1}$$\times$(gaussian$_{\small 2}$ + powerlaw$_{\small 3}$) \\
                   &                      &         &        	&   & + phabs$_{\small 4}$$\times$bbodyrad$_{\small 5}$\\
 
 \hline

 IGR J18027$-$2016   &  0745060401          &   E    &   22150     &	& phabs$_{\small 1}$$\times$(powerlaw$_{\small 2}$ + gaussian$_{\small 3}$  \\
                   &                      &        &     	&   & + gaussian$_{\small 4}$)\\
                     &                    &        &     	&   & \\
                 
 IGR J18027$-$2016   &  0745060401          &   OOE  &   4347  	&   & phabs$_{\small 1}$$\times$powerlaw$_{\small 2}$  \\
 \hline

 IGR J17252$-$3616   &  0405640201          &   E    &  19190  	&   &phabs$_{\small 1}$$\times$(powerlaw$_{\small 2}$  + gaussian$_{\small 3}$  \\
                   &                      &        &     	&   &+ gaussian$_{\small 4}$) + phabs$_{\small 5}$$\times$bbodyrad$_{\small 6}$  \\
 \hline

 IGR J17252$-$3616   &  0405640601          &   E     &  10620  	&   & phabs$_{\small 1}$$\times$(powerlaw$_{\small 2}$ + gaussian$_{\small 3}$)  \\
                   &                      &         &    	&    & + phabs$_{\small 4}$$\times$bbodyrad$_{\small 5}$ \\
 \hline

 IGR J17252$-$3616   &  0405641001          &   E     & 9431   	&   & phabs$_{\small 1}$$\times$(powerlaw$_{\small 2}$ + gaussian$_{\small 3}$)  \\
  \hline
 IGR J16479$-$4514   &  0512180101          &    E   &   18730    &	&  phabs$_{\small 1}$$\times$(gaussian$_{\small 2}$ + gaussian$_{\small 3}$ \\
                   &                      &        &     	&  &  + powerlaw$_{\small 4}$)  + phabs$_{\small 5}$$\times$bbodyrad$_{\small 6}$\\
 \hline

 IGR J16418$-$4532   &  0679810101          &   E    &  16990  	&   & phabs$_{\small 1}$$\times$(powerlaw$_{\small 2}$\\
                   &                      &        &     	&   &+ gaussian$_{\small 3}$ + gaussian$_{\small 4}$) \\
  \hline  
 \end{tabular}
\label{models}
 \end{table*}
%
%
\section{Discussion}
During the X-ray eclipses, direct emission from the compact object is blocked by the companion and the observed X-rays
during eclipses are the reprocessed emission of the primary X-rays from the surrounding medium. For HMXBs, the main 
reprocessing agent is the stellar wind of the companion and any structures in the wind, if present. 
The best fit models for all the sources have been given in Table \ref{models}.
Among the multiple
spectral components, the continuum is expected to be strongly suppressed during the eclipses and the emission lines
which are often produced in a larger region in the stellar wind are expected to be suppressed to a lesser extent.
The eclipse spectrum is therefore expected to show larger equivalent width and better detectability of the 
emission lines compared to the out-of-eclipse spectrum. However depending upon the distribution of the material 
around the compact object and the wind density, this effect can be different in different sources. 
To investigate the phenomena of X-ray reprocessing in the stellar wind in a large number of HMXBs, and 
to find if there are significant system to system differences, we have analyzed 13 eclipse observations of 9 HMXBs  with the 
out-of-eclipse observations, whenever available along with the eclipse phases in the same observation.
We then compare the eclipse spectra of these sources, and also compare the eclipse and out-of-eclipse spectra whenever available.
We have found some similarities and significant differences in the reprocessing properties of the HMXB systems we have analyzed.
All these HMXBs have supergiant companions with stellar mass in the range of $\sim$16M$_{\odot}$ to 52M$_{\odot}$, radius 
in the range of 8R$_{\odot}$ to 26R$_{\odot}$ and orbital period of binaries in the range of 1.41 days 
to 9.74 days. Accretion in three of these systems, LMC X-4, Cen X-3, 
and SMC X-1 are at least in parts due to Roche Lobe overflow \citep{1978Savonije, 2007Meer}. 
\par
A summary of the comparison of various aspects of the X-ray 
reprocessing in these HMXBs is as follows: Equivalent width of the lowly ionized (or neutral) 
Fe K$_{\alpha}$ emission line during eclipse is found 
to have a wide range, from $\sim$126$^{+18}_{-18}$ eV (in SMC X-1, OB ID: 0011450101) to 2695$^{+415}_{-622}$ eV 
(in IGR J17252$-$3616, OB ID: 0405640201)
i.e. a factor of $\textgreater$20 larger in
the later source. Ratio of the equivalent width of these Fe K$_{\alpha}$ emission line during the eclipse 
and the out-of-eclipse phases also have a large range, from $\sim$0.55 (in Cen X-3) to $\sim$10 (in 4U 1538$-$522). 
We have compared the best fit models in the energy range of 0.3-10 keV and 3-10 keV. We have found consistent
value of the equivalent widths of the Fe lines in eclipse and out-of-eclipse phases. So the error quoted for the
equivalent widths are statistical errors only, these do not depend on the choice of the continuum.
Flux ratio of the eclipse to out-of-eclipse
spectra shows a wide range, from $\sim$8 (in 4U 1700$-$377, OB ID: 0083280401) to $\sim$237 (in LMC X-4, OB ID: 0142800101), which differs
by a factor of $\sim$30. Even in the same source LMC X-4, the ratio of flux in the out-of-eclipse and the eclipse phase varies as 
much as by a factor of $\sim$3 (from $\sim$237 to $\sim$86) in different observations.
Here we discuss the results of some individual sources.

\subsection{Cen X-3:}
  In Cen X-3 (OB ID: 0111010101), three Fe emission lines i.e. 6.41 keV
  (fluorescent Fe K$_{\alpha}$) and 6.66 keV (Fe XXV) and 6.94 keV (Fe XXVI) emission lines have been detected. 
 Equivalent width of Fe XXV  and Fe XXVI emission lines have 
 been observed to be higher in the eclipse phase compared to the out-of-eclipse as expected. 
 The equivalent width of the Fe emission lines in this source during eclipse is  smaller than
 for other SgHMXBs with similar line of sight
equivalent Hydrogen column density (N$_{H}$). \cite{1996Ebisawa} and \cite{2012naik} 
have also found lower equivalent width of 
Fe emission lines in eclipse compared to out-of-eclipse with \textit{ASCA} and \textit{XMM-Newton} respectively. The wind in Cen X-3
is smooth \citep{2001Wojdowski}. 
A low density
wind in the surrounding region of the compact object can produce low equivalent width, but why the O type supergiant companion
in Cen X-3 has a thinner wind than  O and B type supergiants in other HMXB systems, is not understood.
 The equivalent width of Fe K$_{\alpha}$ emission line
 has been found to be lower in the eclipse phase compared to the out-of-eclipse phase as observed by  \cite{2012naik} and
 \cite{1996Ebisawa}. Fe K$_{\alpha}$ emission line flux increases by a factor of more than 17 as the source 
 comes out of the eclipse while FeXXV and FeXXVI emission line fluxes increase by a factor of $\sim$5.5 and $\sim$5.8
 respectively. Similar result was obtained by \cite{2012naik}.
 These observations confirm that most of the fluorescent Fe K$_{\alpha}$ 
 line emitting region were closer to the
 source and distributed in a region of size comparable or smaller than the radius of the companion star from the compact 
 object. Ionization state of the Fe atoms are expected to be high near the 
  compact object because of its intense X-ray emission, but here we see most of the low inonized Fe atoms are closer to the 
  compact object compared to the highly ionized Fe aoms.
  This is possible if the 
  Fe atoms closer to the source are in very dense optically thick accretion disc or in dense accretion stream. 
  High density of the disc or the stream keeps
the Fe atoms lowly ionized in spite of them  being near
to the source. 
\subsection{LMC X-4:}
\cite{2004naikandPaul} have observed the source with \textit{BeppoSAX} in low and high 
states of its superorbital period. The decrease of the continuum flux was found to be more by a factor of 
5 compared to the decrease of 6.4 keV Fe line flux in the low state, which they suggest is caused
by a different origin for the continuum and the Fe line emission. The smaller value of Fe line flux and higher value of
Fe line equivalent width in the low state indicates that a major part of the line emission is from
a region comparable to or smaller than the size of the obscuring material (most probably part of the 
accretion disk) and also some part of the emission line comes from an extended region.
In the low state they have not found any significant increase of the line of sight column density, 
indicating that the continuum X-ray emission comes from an extended region.
The equivalent width of fluorescent Fe K$_{\alpha}$ emission line 
have been detected during low and high state of the source with 1260 eV and 240 eV respectively.
\cite{2009Neilsen} found variable Doppler shift of the Fe lines. They suggest that the origin 
of the Fe line is the inner accretion disk warp. The evolution of the other emission lines suggests
precession of the accretion disk.
\par
During eclipse in the first pn observation of LMC X-4 (OB ID: 0142800101) fluorescent
Fe K$_{\alpha}$ emission line 
has been detected with high equivalent width (597$^{+171}_{-256}$ eV),
but this line is absent in the out-of-eclipse phase. 
 The large increase  ($\textgreater$237 times) of the total flux in the 
 out-of-eclipse phase perhaps strongly suppressed this emission line which makes 
 it undetectable during the out-of-eclipse phase.
 Fe XXV emission line has been detected during the out-of-eclipse phase but it is absent
 in the eclipse phase, which
 indicates most of the highly ionized wind material which emits Fe XXV photons is distributed 
 in a region less than
 the distance comparable to the radius of the companion from the  compact object. 
  In the second observation of LMC X-4 (OB ID: 0203500201)
  during eclipse, the  detection of fluorescent Fe K$_{\alpha}$ emission line
  is marginal. However we can not rule out this line as the statistics of the eclipse spectrum of this 
  observation is very poor. During out-of-eclipse phase in the second observation fluorescent
  Fe K$_{\alpha}$ emission line is comparatively stronger. This indicates most of the lowly 
  ionized Fe atoms were closer to the source
 during this observation. When the source came out of the eclipse, these Fe atoms showed their 
 presence through the 6.38 keV emission 
 line during the out-of-eclipse phase. This can happen only if these Fe atoms are in the  dense 
 accretion disc and/or 
 accretion stream during this observation. 
   Highly ionized 6.99 keV Fe XXVI emission line has also been detected during 
 out-of-eclipse phase of the second observation, but it is insignificant in the eclipse phase, which possibly 
 indicates that the  origin of this emission
 line is very near to the compact object. However due to poor statistics 
 of the eclipse spectrum we can not make any definitive 
 comment about the scenario. 
  Highly ionized Fe XXVI emission line has been detected in LMC X-4 for the first time (OB ID: 0203500201). 
   None of the earlier observations have detected this emission line in this source. We checked for the superorbital phase of
   the source during the out-of-eclipse phase of this observation and found it in the bright phase, 
   while the other observation (OB ID: 0142800101) of this source shows a low superorbital phase. 
 During eclipse phase in the first observation of LMC X-4, 
 the total flux (0.3-10.0 keV) is lowered  by a 
 large factor ($\textgreater$237) from the out-of-eclipse phase, while in the second observation 
 the flux ratio between the out-of-eclipse to eclipse phase is $\sim$86.  
 This signifies a difference in the density and/or spatial distribution of the reprocessing material between these two
 observations.
 During eclipse in the first observation lowly ionized (or neutral) fluorescent
 6.36 keV Fe K$_{\alpha}$ emission line has been detected, while in the
  second observation during eclipse phase the detection of this line is insignificant.
  In the first observation during the out-of-eclipse phase amongst Fe emission lines 
  only highly ionized 6.60 keV Fe XXV emission line has been detected, in the second 
  observation during out-of-eclipse phase,   
   6.38 keV  Fe K$_{\alpha}$ emission line and highly ionized 6.99 keV  XXVI emission line have 
 been detected.
 These indicate higher ionization of matter
 near the compact object during the second observation compared to the first observation.
 Lowly ionized (or neutral) 6.38 keV Fe K$_{\alpha}$ and highly ionized Fe XXVI emission 
 during the out-of-eclipse phase of the second observation indicates a mixture of hot and 
 cold wind near the compact object.
 Lowly ionized Fe atoms could possibly be distributed in optically
 thick accretion disk or in the dense accretion stream closer
 to the source.
   The high density of the stream makes them lowly ionized in spite 
   of these Fe atoms being near to the source. 
  \subsection{4U 1700$-$377:}
In all the eclipse, eclipse egress and low flux spectra spectra of 4U 1700$-$377
\cite{2005van_der_Meer} have found a $\sim$6.4 keV Fe emission line, they suggest a clumpy  
wind near the compact object as the origin of this line.
The detection of many strong emission
lines during eclipse indicates that these lines are originating from an extended region.
Also the intensity of some of the emission lines with very high values of their
ionization parameters increase towards egresss, which indicates that the
extension of the ionized plasma around the central source is not very large either.
\par
  Two observations of 4U 1700$-$377 were separated by a span of more than 8 years.
   In the first observation of 4U 1700$-$377 (OB ID:  0083280401) lowly ionized fluorescent
 Fe K$_{\alpha}$ emission line has been detected in both the eclipse and  out-of-eclipse phases
 (6.37 keV and 6.44 keV respectively) which are consistent with energy 6.4 keV.
 The equivalent width of this Fe emission line
 during eclipse is found to be much higher as compared to the out-of-eclipse phase,
as expected. The emission flux and the equivalent width of this
 Fe K$_{\alpha}$ emission line detected during eclipse is found to be higher 
  compared to the same in the other observation (OB ID: 0600950101) during eclipse. 
 This signifies greater reprocessing 
 by lowly ionized Fe atoms during this observation compared to the later one.
 Non-detection of Fe XXV and Fe XXVI emission lines in this observation indicates
 comparatively colder wind during this observation.
 Many low energy emission lines have been detected in the eclipse spectrum of the second observation but were 
 not detected in the eclipse phase of the first observation, which could be due to limited statistics.  
  \par
  The second observation of 4U 1700$-$377 (OB ID: 0600950101) covers only eclipse and in this phase 
   fluorescent Fe K$_{\alpha}$ and highly ionized Fe XXV  emission lines have been detected along 
   with other low energy emission lines and an emission line of 
 energy 7.05 keV. This 7.05 keV  emission line could be Fe K$_{\beta}$ line.
 The ratio of the flux of the Fe K$_{\beta}$  to Fe K$_{\alpha}$ emission line is 0.18($\eta$). This indicates
 ionized medium surrounding the compact object($\eta$ $\leq$0.125 for Fe neutral gas \citealt{1993Kaastra}), 
 which is consistent with the detection of the highly ionized Fe
 K$_{\alpha}$ (Fe XXV) emission line. The equivalent width of fluorescent Fe K$_{\alpha}$ emission line is very high (1061$^{+13}_{-13}$ eV)
 and that of the Fe XXV emission line is quite low (22$^{+4}_{-3}$ eV) in this observation. This indicates an ionized medium
 most of the Fe atoms lower ionized than Fe X  
 are distributed at least at a distance comparable to the radius of the companion star from the compact object
 during this observation.
 \subsection{Other sources}
   Pulse phase resolved spectra of SMC X-1 with \textit{BeppoSAX}
  show a phase shift of the pulsating soft component compared with that of the hard powerlaw component \citep{2004Naik}. 
  This suggests different origin of the soft and hard X-rays.
  \par
   In SMC X-1 a weaker lowly fluorescent 6.38 keV Fe K$_{\alpha}$ emission
   line has been detected with EPIC pn during eclipse with an equivalent width of 126$^{+18}_{-18}$ eV, which is the 
   lowest equivalent width of the fluorescent Fe K$_{\alpha}$ emission line
   amongst all the eclipsing sgHMXBs reported in this work. The detection of
   this line is marginal in the out-of-eclipse phase (with an upper limit of equivalent width of $\sim$90 eV).
   Inspite of having good statistics in the eclipse spectra,
   a weak detection of fluorescent Fe K$_{\alpha}$ emission line could be due to low metalicity of the system 
     \par
     In the eclipse spectrum of 4U 1538$-$522 with \textit{XMM-Newton}, \cite{2011Roca} have found 6.4 keV 
     Fe K$_{\alpha}$ fluorescent and 6.6 keV Fe XXV emission line. They also detected other
     emission lines with wide range of ionization states. This broad range suggests that the
     emiiting region is either wide or the emitting material have large range of densities.
    The coexistence of variety of recombination and fluorescent emission lines has
    also been observed in Vela X-1 during eclipse with \textit{ASCA} and \textit{Chandra} \citep{1999Sako,2002Schulz}, 
    which suggests an inhomogeneous wind structure with cool, dense clumps scattered in highly ionized plasma.
             \par
     In 4U 1538$-$522, the fluorescent Fe K$_{\alpha}$ emission line has been observed in both the 
     eclipse and the out-of-eclipse phase with greater equivalent width during eclipse phase as expected. 
    As pointed out in section 2.2.1, the power-law component in this source is heavily absorbed with a column density of 
  17.16$^{+1.17}_{-1.12}$ $\times$ 10$^{22}$ cm$^{-2}$. A black body component should not be detectable, if it 
  has same column density. We have therefore, fitted with soft excess with 
  a different column density. The line of sight Hydrogen column density for the soft excess is less than
  that for the powerlaw by a large factor
  ($\textgreater$30), which indicates that soft X-ray emissions are coming from a different region away
  from the neutron star.
    Again nearly equal
  soft X-ray flux (0.3-2.0 keV) in the eclipse and 
  the out-of-eclipse spectra indicates the origin of the soft X-rays  far away (at least further away 
  than the size of the companion)
  from the compact object and is not blocked during eclipse.   
  \par
  \cite{2005Hill} have obtained a smaller value of the Fe K$_{\alpha}$ emission line 
  equivalent width for IGR J18027$-$2016 in off and on pulse (40 eV and 25 eV respectively).} We detected
  this Fe emission line with high equivalent width during eclipse (445$^{+254}_{-191}$ eV) in IGR J18027$-$2016 
  but has not been detected in the out-of-eclipse phase, which is similar 
  to the observation of SMC X-1 except that the equivalent width in case of SMC X-1 during eclipse phase was
  found to be low (126$^{+18}_{-18}$ eV). 
    \par
  The observations of IGR J16479$-$4514, IGR J16418$-$4532 and the three observations of 
  IGR J17252$-$3616 covered only the eclipse. In these observations low
  ionized (or neutral) Fe K$_{\alpha}$ emission lines have been detected with large equivalent widths. 
  In the first observation (OB ID: 0405640201) 
  of IGR J17252$-$3616 the equivalent width is found to be largest (2695$^{+415}_{-622}$ eV) amongst all the observations 
  analyzed in this work. 
 Varying equivalent width (921--2695 eV) and flux (0.04--0.18 photons cm$^{-2}$ sec$^{-1}$) of Fe K$_{\alpha}$ 
  emission line in 3 observations
  of IGR J17252$-$3616 indicate a change in the distribution and density of
 lowly ionized Fe atoms surrounding the compact object at least at a distance equal to the radius of the 
 companion over a period of 1 month. 
 \cite{2011Manousakis} have found spectral variation over the orbital 
 phase of IGR J17252$-$3616 with \textit{XMM-Newton}. During the eclipse they found a drop of the fluorescent
 Fe K$_{\alpha}$ emission line equivalent width by a factor $\textgreater$10, that indicates a cocoon-like 
 wind structure surrounding the compact object. 
For IGR J16479$-$4514,  \cite{2013Sidoli} also have obtained a high 
 equivalent width (5500$^{+4500}_{-3700}$ eV) of fluorescent Fe K$_{\alpha}$ emission line 
 during eclipse with \textit{Suzaku}, while it is in the range of 40-280$^{+160}_{-160}$ eV outside of the eclipse.
Based on \textit{XMM-Newton} data, \cite{2012Sidoli_4532} obtain a high Fe K$_{\alpha}$ emission line 
 equivalent width of 3100 eV during eclipse.
   \par
   In the out-of-eclipse phase of the observation of 4U 1538$-$522, 
   also in the eclipse phase of IGR J16479$-$4514 and in the first two observations of IGR J17252$-$3616, 
   the line of sight equivalent Hydrogen column density for the 
   soft X-ray emission are lower by several factor ($\sim$8-680) than that for the power law. 
   This signifies a different origin for the hard and soft
   X-rays in these sources during the observations mentioned above.
   Detection of a soft excess, even when the power-law component is absorbed by a very large column density
   of material ($\sim$10$^{24}$ cm$^{-2}$) is known in some HMXBs (GX 301-2: \citealt{2014nazma}), and it has been
   attributed to a different origin of the soft component compared to the powerlaw component.
\par   
To study the variation of X-ray reprocessing with the orbital parameters of the HMXBs,
we show the orbital period of the system, mass, radius and mass loss rate of the companion (through wind),   
projected semi major axis of the system as functions of  out-of-eclipse to eclipse flux ratios (Figure \ref{6param_flux-ratio}). 
We have not noticed any correlation of these parameters with the flux ratios. In the two observations of 
LMC X-4, the flux ratios vary by a factor of $\sim$3. In Figure \ref{OOE_NH-eqw} we show a relation between equivalent
width (Eqw) of Fe K$_{\alpha}$ emission line and line of sight equivalent Hydrogen column density (N$_{\rm H}$) obtained
during out-of-eclipse phase. We do not see an evident correlation} between the two parameters. A direct correlation is
expected between Eqw and N$_{\rm H}$ during out-of-eclipse phase, as this line is produced by the fluorescence of X-rays
from the compact object where the surrounding medium causes this fluorescence \citep{1991George}.Thus, in a denser
surrounding medium, more fluorescence is expected to occur.
A direct correlation has been observed between the two parameters by \cite{2010Torrejon}
 and  \cite{2015Garcia}. Out of 7 out-of-eclipse observations we have detected fluorescent Fe K$_{\alpha}$ emission line 
 in 4 observations.} In Figure \ref{OOE_NH-eqw} we also show upper limit of the equivalent width
 of 6.4 keV emission line in other three out-of-eclipse observations. This limited data does not allow a clear conclusion about
 the relationship between Eqw-N$_{\rm H}$ in SgHMXB systems.
 \par
In Figure \ref{NH_flux}, we have plotted N$_{\rm H}$ for the observations which have both the
eclipse and out-of-eclipse data. In this Figure, the 
X axis indicates the observations and the Y axis in the top panel shows N$_{\rm H}$ during eclipse 
(pink boxes) and out-of-eclipse  (blue triangles) phases.
Y axis in the bottom panel gives the value of 0.3-10 keV out-of-eclipse to eclipse flux ratios.
For 3 observations (4U 1700$-$377 OB ID: 0083280401,  4U 1538$-$522, IGR J18027$-$2016) N$_{\rm H}$ 
is larger during eclipse  than out-of-eclipse   
phase. For the other 4 observations, N$_{\rm H}$ is comparable in both the phases.
In Figure \ref{NH_flux_ratio_relation}, we have plotted the out-of-eclipse to
eclipse flux ratios with the ratio of line of sight equivalent Hydrogen column densities (N$_{\rm H}$)
during out-of-eclipse and eclipse of the HMXBs, for which both the out-of-eclipse and eclipse
data were available. We excluded LMC X-4 from this plot, as the N$_{\rm H}$ could not be constrained for this source.
The sources that show large flux ratio between out-of-eclipse and eclipse spectra 
(SMC X-1, IGR J18027$-$2016, and 4U 1538$-$522) probably have less dense wind environment compared to the 
two source Cen X-3 and 4U 1700$-$47, as the large out-of-eclipse to eclipse flux ratio indicates
lesser reprocessing. Within each type, the sources with low ratio of the column density
in the out-of-eclipse and eclipse spectra (SMC X-1 and Cen X-3) probably have 
more isotropic wind pattern compared to the sources for which the column density is much larger in the 
out-of-eclipse spectra (4U 1700$-$37, IGR J18027$-$2016, and 4U 1538$-$522). In the later sources, the supergiant
companion might have stronger wind outflow in the equatorial (or near equatorial) plane (as shown in Figure 
\ref{dense_flow}), the companion could also have a trailing wind as described by \cite{1994Blondin} as shadow wind.
These features may explain the differences in column density in the eclipse and out-of-eclipse phases.
\par
Comparison of the fluxes in (0.3-2.0) keV energy range in eclipse and out-of eclipse phases indicate source of the soft
excess is near the compact object in our sample of HMXBs except in 4U 1538$-$522. In 4U 1538$-$522 the soft excess in 
both the phases are comparable, which indicates that the soft excess must be originating far away from the system. 
Spectrum also softens during eclipse, perhaps the soft excess is due to dust scattering in the interstellar medium (ISM) 
along the line of sight. \cite{2001Robba} has also observed softening of the spectrum during total eclipse compared to the 
out-of-eclipse phase in 4U 1538$-$522, they suggest dust-scattering of the direct photon as the origin of the soft excess
in the source. \cite{2006Audley} have also found evidence of dust hallow surrounding the eclipsing HMXB OAO 1657$-$415.
\subsection{Comparison of SFXTs and SgHMXBs:} 
SFXTs have a much lower average X-ray luminosity compared to the SgHMXBs and only
become bright during short flares. The compact objects in SFXTs are most probably
neutron stars \citep{2005zand}.
    The SFXTs show lower equivalent width of Fe K$_\alpha$ emission line compared to 
  SgHMXBs ouside eclipse \citep{2018Pradhan}
  with \textit{Suzaku} and \textit{XMM-Newton} data. Their findings in a large sample of both the 
  systems suggest that the compact object 
  in the SFXT systems are surrounded by winds less dense than that in SgHMXBs. The compact object
  in SFXTs is inhibited 
  from accretion from its companion most of the time  \citep{2006Negueruela, 2005Sguera},
  the low intensity X-ray emission is insufficient to reduce the 
  speed of the supergiant's radiatively driven wind effectively, hence most of the wind pass by the 
  compact object and lost from the SFXT systems and the
  compact objects in SFXTs are surrounded by less dense winds. 
  \cite{2014Maccarone}
  suggests that some 
  SFXTs might have eccentric orbit,
so the compact object spends most of the time away from its supergiant companion
 and hence surrounded by less dense matter.
 But as a counterexample, indications of a circular orbit has been found in SFXT IGR J16479$-$4514 which 
 has an orbital period of $\sim$3 days \citep{2009Jain, 2013Sidoli}.
 Considering a spherical wind outflow from the supergiant 
 star and from the scattered and direct emission respectively during eclipse and out of eclipse 
 phases an expected luminosity of these sources can be estimated, but this is
 two orders higher than the observed luminosity \citep{2017nunez, 2013Sidoli}.
  The usual low luminosity in the SFXT systems 
can be explained if the neutron star remains
  in supersonic propeller regime most of the time. This can lower the luminosity by a factor
  of 100--1000 compared to the case of direct 
  accretion for neutron stars with spin period in the range of 10--100 s and magnetic field 
  with $\mu_{30}$ = 0.1--1  \citep{2017nunez}.
  If the interaction of the supergiant's
  wind with the compact object in these 
  systems is very different than that in SgHMXBs, that also could make the compact object's environment less dense,   
  which produces low equivalent widths of Fe emission line \citep{2018Pradhan}.
    \par
  However the two SFXTs (IGR J16479$-$4514, IGR J16418$-$4532) in our sample have Fe line equivalent widths large and comparable
  to the SgHMXBs during eclipses. IGR J16479$-$4514 and IGR J16418$-$4532 have equivalent widths of 803$^{+89}_{-89}$ eV and 1358$^{+453}_{-453}$
   eV respectively for the fluorescent Fe K$_\alpha$ emission line
   during eclipses, while the equivalent width of this line for the SgHMXBs is in the range of
   (126--2695 eV). The equivalent width of the Fe XXVI emission line are 576$^{+345}_{-230}$ eV and 1911$^{+1146}_{-764}$ eV, respectively
   in the above two SFXTs respectively.  The line of sight Hydrogen column density (N$_{\rm H}$) in these SFXTs
   are like other SgHMXBs in our sample. 
   During eclipse, both the continuum and the Fe line are reprocessed emission from the same wind material, so a comparable 
   equivalent width of the Fe line in SgHMXBs and SFXTs during eclipse only indicate similar Fe abundance in the two kinds of
   systems. A lower Fe abundance can possibly explain a lower line equivalent width in SFXTs outside the eclipse (as found by 
   \citealt{2018Pradhan}), but such a possibility is ruled out with the current result.
      We need large samples with multiple observations, during eclipses to interpret the relation between the
   fluorescent Fe K$_\alpha$ emission line equivalent width with luminosity and N$_{\rm H}$, 
   also to investigate the interaction of the compact object with the supergiant's wind 
  and hence wind distributions surrounding the compact objects in SFXTs. 
  \section{Conclusion} 
We have found ample diversity in the X-ray reprocessing characteristics in high mass X-ray binaries.
The out-of-eclipse to eclipse flux ratio has been found to be in the range of $\sim$(8--237), which implies 
significantly dynamic wind structure surrounding the compact object in HMXBs. Even in the same source at
different epochs the variation is quite large (86--237 in LMC X-4).

The equivalent widths of Fe emission lines found in SFXTs are large during eclipse, similar to that in
SgHMXBs. The equivalent width of the fluorescent Fe K$_\alpha$ emission line found in the SFXTs 
IGR J16479$-$4514 and IGR J16418$-$4532 are 803$^{+89}_{-89}$ eV and 1358$^{+453}_{-453}$ eV, respectively. Whereas the equivalent 
width of the Fe XXVI emission line have been found to be 576$^{+345}_{-230}$ eV and 1911$^{+1146}_{-764}$ eV respectively
in these two SFXTs. Which implies a similar Fe rich medium in SFXTs like in SgHMXBs.

Cen X-3 is one exception for which the equivalent width of the Fe K$_\alpha$ emission line 
is lower during eclipse compared to the out-of-eclipse phase, this indicates an Fe rich, dense 
accretion disk or accretion stream near the compact object.

In 4U 1538$-$522 the soft X-ray emission flux is nearly same in out-of-eclipse and eclipse phase, i.e. the soft
X-ray emission is not blocked during eclipse of the compact object by the supergiant companion, which
clearly indicates a different origin for the hard and soft X-rays. Perhaps dust scattering of the direct photons
in ISM, far away from the source are the origin of this soft excess. 

Some systems show comparable wind density near and far away from the compact object (Cen X-3, LMC X-4,
SMC X-1). There are some indications of equatorial or near equatorial dense wind outflows from the supergiant
in three different observations (IGR J18027$-$2016, 4U 1538$-$522 and first observation of 4U 1700$-$377).\\
\\
\textit{\Large{Acknowledgements:}} \\
 \\
This work has made use of archival
data obtained from \textit{XMM-Newton} Science Archive (XSA) provided by the European Space Agency 
(ESA). We have also used the public light curves 
from the \textit{Swift}-BAT site. 
%
%
\begin{table*}
\caption{Line of sight Hydrogen column density (N$_{\rm H}$), photon index ($\Gamma$) and blackbody temparature (T$_{\rm BB}$),
$\chi^{2}$/DOF and total flux in the energy range of (0.3-10.0) keV of 9 eclipsing HMXB systems during eclipse (E) and out-of-eclipse (OOE) phases}
\centering
\resizebox{1.0\textwidth}{!}
{\begin{tabular}{lcccccrr}
 \hline
 
 Source  	  & Observation  	 & Phase              & N$_{\rm H}$            	 & Photon        	        & T$_{\rm BB}$ 		 	& $\chi^{2}$ (DOF) &  Total Flux      \\
		  & ID          	 &          	      &		    		 & Index         	        &           		 	& 	                             \\
		  &             	 &       	      & (10$^{22}$               & ($\Gamma$)                   & (keV)                  	& 	          &  (10$^{-11}$ ergs \\
                  &             	 &                    & cm$^{-2}$)               &                              &                        	& 	          &   cm$^{-2}$ sec$^{-1}$) \\
\hline
&& && && &\\
Cen X-3          &   0111010101    	 & E          	      &   1.50$^{+0.10}_{-0.11}$  & 0.71$^{+0.02}_{-0.06}$      &  0.14$^{+0.01}_{-0.01}$       & 197.73 (127)    &	  4.47                  \\
Cen X-3          &   0111010101 	 & OOE 	      	      &   2.07$^{+0.18}_{-0.16}$  & 0.51$^{+0.05}_{-0.02}$      &  0.08$^{+0.01}_{-0.01}$       & 207.39 (137)     &	  43.74                 \\

&& && && &\\

 LMC X-4         &  0142800101   &  E      		     & 0.06 (frozen)             &  0.0$^{+0.3}_{-0.3}$       & 0.22$^{+0.03}_{-0.03}$ 	& 38.83 (37)	  &        0.11            \\
 LMC X-4         &  0142800101   &  OOE  		     & 0.06 (frozen)             &  0.74$^{+0.01}_{-0.01}$    & 0.044$^{+0.005}_{-0.005}$    & 1129.22 (718)	  &        26.09   \\

 && && &&  & \\
 
    LMC X-4       &  0203500201 &  E       		    & 0.06 (frozen)    		 & 0.08$^{+0.20}_{-0.21}$      &   0.18$^{+0.02}_{-0.01}$      & 103.01 (94)       &   0.18       \\
    LMC X-4       &  0203500201 &  OOE  		    & 0.06 (frozen)	 	& 0.88$^{+0.04}_{-0.04}$       &   -     		        & 151.81 (145)      &   15.48             \\

 && && &&  & \\

    SMC X-1       &  0011450101  &  E     		   &  0.03$^{+0.02}_{-0.02}$      & -0.56$^{+0.23}_{-0.25}$         &  0.23$^{+0.02}_{-0.01}$           & 181.73 (147)	   &   0.52        \\
    SMC X-1       &  0011450101  &   OOE 		   &  0.06$^{+0.03}_{-0.03}$      & 0.20$^{+0.17}_{-0.17}$          & 0.21$^{+0.02}_{-0.02}$            & 198.21 (156)   &	39.84            \\

 && && &&  & \\

  4U 1700$-$377     &  0083280401  &  E  		   & 0.62$^{+1.10}_{-0.62}$         & -0.41$^{+0.26}_{-0.24}$           & -                          & 88.60 (71)	   & 4.28          \\
  4U 1700$-$377     &  0083280401  &  OOE              & 23.17$^{+19.36}_{-6.36}$       & -0.17$^{+0.52}_{-0.55}$           & -  			     & 23.96 (18)	   & 34.28           \\

   && && &&   &\\
 
     4U 1700$-$377     &  0600950101  & E     		   &  1.01$^{+0.32}_{-0.43}$      &-1.34$^{+0.14}_{-0.14}$               &  0.08$^{+0.04}_{-0.06}$           & 195.06 (133)	   & 3.19            \\

 && && &&  & \\
 
   4U 1538$-$522     &  0152780201  &    E       		   & 0.36$^{+0.08}_{-0.05}$         & 0.22$^{+0.06}_{-0.06}$           &  -                         & 178.19 (130)   &  0.46    \\
   4U 1538$-$522     &  0152780201  &   OOE  		   & 17.16$^{+1.17}_{-1.12}$        & 0.71$^{+0.08}_{-0.08}$           &  1.53$^{+0.88}_{-0.42}$   	    & 125.90 (126)   &  13.93          \\

 &&  && &&   &\\
 
   IGR J18027$-$2016  &  0745060401  &    E       	   &  0.36$^{+0.53}_{-0.36}$         & -0.22$^{+0.24}_{-0.23}$          &   -          		    & 49.28 (41)      &  0.08           \\
   IGR J18027$-$2016  &  0745060401  &   OOE   		   & 5.90$^{+0.40}_{-0.37}$         & 0.71$^{+0.08}_{-0.08}$           & -           		    & 110.04 (110)    &    3.47   \\

 && && &&    &\\

  IGR J17252$-$3616 &   0405640201  &  E        		  &9.34$^{+137.59}_{-6.46}$    		& 0.24$^{+0.83}_{-0.75}$             &  0.19$^{+4.92}_{-0.10}$           & 6.23 (7)   &	 0.04          \\

  && && &&   & \\
 
  IGR J17252$-$3616 &   0405640601   &  E         	  & 18.17$^{+24.96}_{-12.15}$    & 0.81$^{+4.92}_{-1.17}$             &   1.07$^{+0.01}_{-0.001}$         &7.66 (8)   &	  0.05          \\

  && && &&   & \\
 
  IGR J17252$-$3616 &   0405641001  & E      		  & 13.48$^{+43.32}_{-13.48}$               & -0.52$^{+4.35}_{-2.47}$            &   -          &   3.50 (6)   &  0.03       \\
&& && && &\\

   IGR J16479$-$4514 &  0512180101  &  E       		   &  0.58 (frozen)              & 1.28$^{+0.37}_{-0.34}$            & 0.07$^{+0.39}_{-0.09}$            & 	44.67 (36)     &    0.08      \\

 &&  && &&   &\\
 
 IGR J16418$-$4532 &  0679810101    &  E   & 0.84$^{+1.43}_{-0.84}$          	 & 0.40$^{+0.83}_{-0.45}$      	&   -          & 2.13 (5)	   &     0.03      \\

 &&  && &&  &  \\

\hline
\end{tabular}}
   \label{NH}
\end{table*}
\begin{turnpage}
\begin{table*}
\begin{center}
\caption{\small{Fe K$_{\alpha}$ (fluorescent), Fe XXV and Fe XXVI emission line energy, flux and 
equivalent width for 9 HMXBs during eclipse (E) and out-of-eclipse (OOE) phases. Line fluxes 
are given in units of 10$^{-4}$ photons cm$^{-2}$ sec$^{-1}$.}}
\resizebox{1.3\textwidth}{!}
{\begin{tabular}{lc cc cc cc cc cc}
\hline
Source  & Observation        &State  &Fe                        & Fe            	  & Fe                            &Fe                        & Fe  	               & Fe               	      & Fe     			 & Fe      			& Fe  		\\
	 & ID                &	     &K$_{\alpha}$              & K$_{\alpha}$  	  & K$_{\alpha}$                  &XXV                       & XXV   	               & XXV              	      & XXVI   			 & XXVI	 			& XXVI	        \\
	 &                   &	     & line                     & line          	  & line                          & line                     & line                    & line              	      & line 			 & line   			& line          \\
	 &                   &	     &energy                    & flux   	          & eqv.                          &energy   		     & flux 	               & eqv.              	      &energy			 & flux    			& eqv.		\\
	 &                   &	     &	          	        &    	      		  & width                         &	      		     & 	                       & width            	      &	      			 & 	  			& width	        \\
 	 &                   &	     &(keV)	                &    	                  & (eV)                          &(keV)    		     & 	                       & (eV)                         & (keV)	          	 &    	                        & (eV) 			        \\
 	 
 \hline
    && && && && && & \\

Cen X-3 & 0111010101         &E     &6.41$^{+0.05}_{-0.03}$    & 0.88$^{+0.08}_{-0.09}$  & 104$^{+9}_{-11}$      & 6.66$^{+0.01}_{-0.01}$  &2.12$^{+0.09}_{-0.1}$    &246$^{+10}_{-12}$    & 6.94$^{+0.01}_{-0.01}$    &1.65$^{+0.09}_{-0.09}$	&209$^{+11}_{-11}$  \\
Cen X-3 & 0111010101         &OOE   &6.41$^{+0.01}_{-0.01}$    & 15.01$^{+0.93}_{-0.66}$ & 189$^{+12}_{-8}$       & 6.68$^{+0.01}_{-0.01}$  &11.64$^{+0.45}_{-0.58}$  & 131$^{+5}_{-7}$     & 6.98$^{+0.01}_{-0.01}$    &9.58$^{+0.49}_{-0.50}$       &139$^{+7}_{-7}$  \\ 
 && && && && && & \\

 LMC X-4 & 0142800101         &E     &6.36$^{+0.26}_{-0.11}$    &0.07$^{+0.02}_{-0.03}$   & 597$^{+171}_{-256}$   &- 	                     &-	                       &-			      &- 	                  &-	  			&- \\
 LMC X-4 & 0142800101         &OOE   & -      	       	  & -		            & -                              & 6.60$^{+0.04}_{-0.03}$  &4.03$^{+0.28}_{-0.26}$   &166$^{+12}_{-11}$      &- 	                  &-	  			&- \\
    && && && && && & \\
 LMC X-4 & 0203500201         &OOE   &6.38$^{+0.04}_{-0.04}$    &1.55$^{+0.03}_{-0.03}$	  & 117$^{+2}_{-2}$       &- 	                     &-	         	       &-                      	      &6.99$^{+0.05}_{-0.05}$     &2.39$^{+0.05}_{-0.04}$ 	&195$^{+4}_{-3}$ \\
    && && && && && & \\

 SMC X-1 & 0011450101         &E     &6.38$^{+0.06}_{-0.06}$    &0.07$^{+0.01}_{-0.01}$  &  126$^{+18}_{-18}$   &-                        &-	       		       &-			      &- 	                  &-	 			&- \\
 SMC X-1 & 0011450101         &OOE   &-                         &-    	                  &- 		                   &-                        &-	       		       &-			      &- 	                  &-	 			&- \\
   && && && && && & \\

 4U 1700$-$377 & 0083280401     & E    &6.37$^{+0.02}_{-0.02}$    &5.96$^{+0.48}_{-0.49}$   & 1258$^{+101}_{-103}$  &-                        &-	          	       &-	 		      &-                          &-  				&- \\
 4U 1700$-$377 & 0083280401     &OOE   &6.44$^{+0.06}_{-0.06}$    &14.88$^{+5.55}_{-5.43}$  & 191$^{+71}_{-70}$     &-                        &-	        	       &-	  		      &-                          &-  				&- \\
&& && && && && & \\

4U 1700$-$377 & 0600950101     &E     &6.39$^{+0.002}_{-0.002}$  &4.89$^{+0.06}_{-0.06}$   &1061$^{+13}_{-13}$     & 6.68$^{+0.05}_{-0.04}$  &0.26$^{+0.05}_{-0.04}$   &22$^{+4}_{-3}$       &-                          &-  				&- \\
    && && && && && & \\

4U 1538$-$522 & 0152780201     &E     &6.38$^{+0.01}_{-0.01}$    &0.57$^{+0.03}_{-0.03}$   & 792$^{+42}_{-42}$     &6.69$^{+0.02}_{-0.02}$   & 0.17$^{+0.02}_{-0.02}$  &145$^{+17}_{-17}$    & 6.95$^{+0.02}_{-0.02}$    &0.13$^{+0.01}_{-0.01}$	&181$^{+14}_{-14}$ \\
4U 1538$-$522 & 0152780201     &OOE   &6.42$^{+0.07}_{-0.07}$    &2.46$^{+0.64}_{-0.57}$   & 79$^{+22}_{-18}$      &-                        &-        		       &-       		      &-		          &-	                        &- \\		
    && && && && && & \\

IGR J18027$-$2016 & 0745060401 &E     &6.41$^{+0.06}_{-0.05}$    &0.07$^{+0.04}_{-0.03}$   & 445$^{+254}_{-191}$  &6.66$^{+0.32}_{-0.36}$    &0.07$^{+0.03}_{-0.04}$  &382$^{+164}_{-218}$			      &		          &				&  \\ 
IGR J18027$-$2016 & 0745060401 &OOE   &-      	                 &-                        &-                     &-                         &-        		      &-			      &-		          &-				&-  \\ 

    && && && && && & \\

IGRJ17252$-$3616 & 0405640201  &E     &6.40$^{+0.03}_{-0.03}$    &0.13$^{+0.02}_{-0.03}$   & 2695$^{+415}_{-622}$  &-                        &-       			&-	 		      &-		          &-				&-  \\ 	
    && && && && && & \\

IGRJ17252$-$3616 & 0405640601  &E     &6.41$^{+0.02}_{-0.02}$    &0.18$^{+0.04}_{-0.03}$   &1831$^{+407}_{-305}$   &-                        &-        			&-			      &-		          &-				&-  \\ 	
    && && && && && & \\
IGRJ17252$-$3616 & 04056401001 &E     &6.40$^{+0.33}_{-0.27}$    &0.04$^{+0.03}_{-0.02}$   & 921$^{+691}_{-460}$   &-                        &-        			&-			      &-		          &-				&-  \\ 	        	        
    && && && && && & \\
IGR J16479$-$4514 & 0512180101 &E     &6.41$^{+0.02}_{-0.02}$   &0.09$^{+0.01}_{-0.01}$    & 803$^{+89}_{-89}$     &-		             &-  	                &-                            & 6.88$^{+0.23}_{-0.12}$    & 0.05$^{+0.26}_{-0.18}$	& 576$^{+345}_{-230}$   \\
    && && && && && & \\ 

IGR J16418$-$4532 & 0679810101 &E     &6.43$^{+0.04}_{-0.04}$    &0.06$^{+0.02}_{-0.02}$   & 1358$^{+453}_{-453}$  &-                        &-        			&-			      & 6.94$^{+0.36}_{-0.27}$	  &0.05$^{+0.03}_{-0.02}$	& 1911$^{+1146}_{-764}$  \\ 	
     && && && && && & \\

\hline
\end{tabular}}
\label{iron}
\end{center}
\end{table*} 
\end{turnpage}

\begin{table*}
 \caption{Ratios of out-of-eclipse to eclipse (OOE/E) fluxes in  0.3-10 keV energy range and equivalent widths of Fe emission lines}
 \centering
 \begin{tabular}{lc rc cc}	
 \hline
 Source 	   & Observation ID   	 &   Flux          & ratio of	        & ratio of	& ratio of	\\
                   &     	      	 &   ratio  	   & eqv width	        & eqv width     & eqv width     \\
                   &     	      	 &  (0.3-10 keV)   & (Fluorescent Fe K$_{\alpha}$)  	& (Fe XXV)      & (Fe XXVI)		\\
                   &     	      	 &  		   &   & 		& 		\\                 
 \hline
 Cen X-3          &  0111010101   	 &    9.79    	   &1.82   	 	& 0.53    	&0.66		\\
 LMC X-4 &  0142800101   	 &  237.18         &-   	 	&-     	 	&-  		\\
 LMC X-4 &  0203500201 	 &   86.00    	   &-   	 	&-     	 	&- 			\\
 SMC X-1         &  0011450101  	 &   76.62   	   &-  	 		&-     	 	&- 			\\
 4U 1700$-$377       &  0083280401  	 &    8.01         &0.15 	 	&-     	 	&-  			\\
 4U 1538$-$522       &  0152780201  	 &   30.28   	   & 0.1  	 	&-     	 	&- 			\\
 IGR J18027$-$2016   &  0745060401  	 &   43.38   	   &-   	 	&-     	 	&- 			\\
 \hline 
 \end{tabular}
 \label{ratios}
 \end{table*}
\begin{table*}
\label{table}
\caption{Emission line energy, flux and  equivalent width other than Fe emission lines. 
Line fluxes are given in units of 10$^{-4}$ photons cm$^{-2}$ sec$^{-1}$.}

\centering
\begin{tabular}{ll ll ll l}
  \hline

 Source& Line   		& Phase    	& Emission   			& Emission  line               	&Line 	                                	&Previous detection\\
 name  & energy	 		&		& line flux	      		& eqv.  width       		&identification	                        	& (keV, line)		\\
 (OB ID)& (keV)			&		&         			&  (eV)                          &	                                	&	\\
 \hline
Cen X-3& 0.98$^{+0.01}_{-0.02}$& E		& 10.5$^{+5.5}_{-3.6}$  	& 96$^{+50}_{-33}$ 	&  Ne X  		                	 & 1.022, Ne X (eclipse) $^{[v1]}$;  \\
 (0111010101)                   &		&				&	      			&    	   	   		&						 & 1.022, Ne X (All phase ranges) $^{[v2]}$ \\ 
        & 			&		&	      			&    	   	   		&	                                	 &			\\ 
        &1.43$^{+0.02}_{-0.02}$	& E		& 0.83$^{+0.27}_{-0.25}$  	& 37$^{+12}_{-11}$	& Mg XII		                	 & 1.472, Mg XII (All phase ranges) $^{[v2]}$ \\
        & 			&		&	      			&    	   	   		&                                                &				\\ 
        &1.98$^{+0.01}_{-0.02}$	& E		& 1.24$^{+0.25}_{-0.20}$   	& 102$^{+20}_{-16}$	&  Si XIV and / P K$_{\alpha1}$, K$_{\alpha2}$	 & 2.00 $\pm$ 0.01, Si XIV (eclipse) $^{[v1]}$\\
        & 			&		&	      			&    	   	   		&                                                &				\\  
        &2.62$^{+0.01}_{-0.02}$	& E		& 0.91$^{+0.15}_{-0.11}$  	& 94$^{+15}_{-11}$	& Cl K$_{\alpha1}$, K$_{\alpha2}$ and /	Si XVI	 & 2.64 $\pm$ 0.02, Si XVI (eclipse) $^{[v1]}$;\\
        & 			&		&	      			&    	   	   		&                                                & 2.621, S XVI (All phase ranges) $^{[v2]}$\\ 
        &  			&		&	      			&    	   	   		&				 		 &\\ 	
        &3.34$^{+0.04}_{-0.04}$	& E		& 0.29$^{+0.08}_{-0.08}$  	& 35$^{+10}_{-10}$	& K K$_{\alpha1}$, K$_{\alpha2}$		 &				- \\
        & 	 		&		&	      			&    	   	   						\\ 
 \hline
                              
        &1.34$^{+0.04}_{-0.06}$	& OOE		& 65.47$^{+24.06}_{-18.12}$	& 345$^{+127}_{-95}$   & Mg K$_{\beta1}$  		                 &          				- \\  
        & 			&		&	      			&    	   	   		&				                 &                                          \\ 
        & 1.79$^{+0.03}_{-0.03}$& OOE		& 12.23$^{+3.27}_{-2.65}$	& 114$^{+30}_{-25}$    & Si  K$_{\alpha1}$, K$_{\alpha2}$, K$_{\beta1}$ &       		1.79 $\pm$ 0.05, Si I and or Si XIII \\
  	&		        &	      	&    	   	   		&				&                                                &                      (pre-eclipse) $^{[v1]}$\\ 
        &2.02$^{+0.01}_{-0.01}$	& OOE		& 5.51$^{+0.89}_{-1.1}$  	& 53$^{+9}_{-11}$ 	&  Si XIV and / P K$_{\alpha1}$, K$_{\alpha2}$ 	 & 			1.99 $\pm$ 0.02, Si XIV (pre-eclipse) $^{[v1]}$; \\
	&		        &	      	&    	   	   		&                               &                                                & 2.006, Si XIV (All phase ranges) $^{[v2]}$\\ 
    	&		        &	      	&    	   	   		&				&		 \\ 
        & 2.64$^{+0.02}_{-0.02}$& OOE		& 2.48$^{+0.39}_{-0.40}$  	& 29$^{+5}_{-5}$	& Cl K$_{\alpha1}$, K$_{\alpha2}$ and /	Si XVI   &			2.64 (fixed), Si XVI (pre-eclipse) $^{[v1]}$; \\ 
	&		        &	      	&    	   	   		&  		                &                                                & 2.621, S XVI (All phase ranges) $^{[v2]}$\\ 
  	&	                &        	&    	   	   		&                               &                                  		 &		 \\ 
        &6.19$^{+0.15}_{-0.19}$	& OOE		& 10.41$^{+2.39}_{-2.80}$  	& 133$^{+31}_{-36}$	&-		                                 &		-\\
  	&		        &		&	      			&    	   	   		&		                                 &			\\ 
\hline
 LMC X-4&0.96$^{+0.01}_{-0.01}$& OOE		& 7.36$^{+0.05}_{-0.05}$ 	& 29$^{+0}_{-0}$	& Cu L$_{\beta}$ /			        & 0.91434 keV (13.560  A$^\circ$), Ne IX He$_{\alpha}$ $^{[w1]}$	\\
 (0142800101)& 			&         	&		                &				&  Zn L$_{\alpha}$ 	      			& 1.01977 keV (12.158  A$^\circ$), Ne X Ly$_{\alpha}$  $^{[w1]}$		\\
   	& 			&         	&				&				&	      					&    	   	   			 							\\
 \hline
                              
 LMC X-4 &0.98$^{+0.03}_{-0.03}$& OOE		& 5.65$^{+2.02}_{-1.54}$  	& 34$^{+12}_{-9}$ 	& Cu L$_{\beta}$ /			        & 0.91434 keV (13.560  A$^\circ$), Ne IX He$_{\alpha}$ $^{[w1]}$	\\
 (0203500201)& 		         &         	&				&	      	                & Zn L$_{\alpha}$	                        & 1.01977 keV (12.158  A$^\circ$), Ne X Ly$_{\alpha}$ $^{[w1]}$		\\
  	      			 &         	&				&	      	                &    	   	   						\\ 
 \hline
 4U 1700--377 &0.81$^{+0.10}_{-0.13}$	& E	& 8.11$^{+32.5}_{-3.95}$  	& 90$^{+359}_{-44}$ 	  & Ne K$_{\alpha}$, O VIII RRC		& 0.86 $\pm$ 0.02, Ne K$_{\alpha}$, O VIII RRC (eclipse)  $^{[x1]}$			\\
 (0600950101) &         	        &	&				&	      			  &    	   	   			&			\\ 
	&1.28$^{+0.05}_{-0.05}$	& E		& 2.05$^{+0.07}_{-0.40}$  	& 1208$^{+41}_{-236}$ 	  & Mg K$_{\alpha1}$, K$_{\alpha2}$, K$_{\beta1}$  & 1.30 $\pm$ 0.01, Mg K$_{\alpha}$, Mg XI He$_{\alpha}$ (eclipse)  $^{[x1]}$		\\
	& 			&         	&				&				  &	      			        &    	   	   						\\ 
	&1.75$^{+0.01}_{-0.01}$	& E		& 0.87$^{+0.14}_{-0.11}$  	& 898$^{+144}_{-113}$	  & Si K$_{\alpha1}$, K$_{\alpha2}$ / Al XIII Ly$_{\alpha}$	& 1.75 $\pm$ 0.01, Si K$_{\alpha}$, Al XIII Ly$_{\alpha}$ (eclipse)  $^{[x1]}$	\\
	& 			&         	&				&				  &	      			        &    	   	   						\\ 
	&1.97$^{+0.03}_{-0.03}$	& E		& 0.16$^{+0.03}_{-0.03}$  	& 139$^{+26}_{-26}$ 	    	  & Si XIV Ly$_{\alpha}$		& 2.00 $\pm$ 0.01, Si XIV Ly$_{\alpha}$ (eclipse)  $^{[x1]}$				\\
 	&			&         	&				&				  &	      			        &    	   	   						\\ 
	&2.34$^{+0.01}_{-0.01}$	& E		& 0.52$^{+0.06}_{-0.05}$  	& 392$^{+45}_{-38}$ 	  & S K$_{\alpha}$, Al XIII RRC		& 2.33 $\pm$ 0.01, S K$_{\alpha}$, Al XIII RRC (eclipse)  $^{[x1]}$			\\
 	&			&         	&				&				  &	      			        &    	   	   						\\ 
	&2.57$^{+0.05}_{-0.06}$	& E		& 0.08$^{+0.02}_{-0.02}$  	& 54$^{+14}_{-14}$ 	  & Cl K$_{\alpha1}$, K$_{\alpha2}$ /	& 2.63 $\pm$ 0.01, S XVI Ly$_{\alpha}$, Si XIV RRC		\\
	& 			&         	&				&				  & S XVI Ly$_{\alpha}$, Si XIV RRC     &  (eclipse)  $^{[x1]}$  	   	   						\\ 
 	&			&         	&				&				  &	      			        &    	   	   						\\ 
 	&2.99$^{+0.01}_{-0.01}$	& E		& 0.21$^{+0.02}_{-0.02}$  	& 115$^{+11}_{-11}$ 	  & Ar K$_{\alpha1}$, K$_{\alpha2}$	& 2.97 $\pm$ 0.01, Ar K$_{\alpha}$ (eclipse)  $^{[x1]}$				\\
 	& 			&         	&				&				  &	      			        &    	   	   						\\ 
 	&3.70$^{+0.01}_{-0.01}$	& E		& 0.25$^{+0.03}_{-0.03}$  	& 102$^{+12}_{-12}$ 	  & Ca K$_{\alpha1}$, K$_{\alpha2}$	& 3.71 $\pm$ 0.01, Ca K$_{\alpha}$ (eclipse)  $^{[x1]}$				\\
 	& 			&         	&				&				  &	      			        &    	   	   						\\ 
 	&4.12$^{+0.05}_{-0.05}$	& E		& 0.08$^{+0.02}_{-0.02}$  	& 29$^{+7}_{-7}$ 	  & Sn L$_{\gamma1}$ / Sb L$_{\beta2}$ /&-								\\
  	&			&         	&				&				  & Xe L$_{\alpha}$ / Sc K$_{\alpha1}$, K$_{\alpha2}$  &    	   	   				\\ 
  	& 			&         	&				&				  &	      			        &    	   	   						\\ 
 	&7.05$^{+0.02}_{-0.01}$	& E		& 0.88$^{+0.10}_{-0.13}$  	& 219$^{+25}_{-32}$ 	  & Fe 	K$_{\beta1}$			&-								\\
 	& 			&         	&				&				  &	      			        &    	   	   						\\ 
 	&7.49$^{+0.02}_{-0.02}$	& E		& 0.29$^{+0.06}_{-0.05}$  	& 71$^{+15}_{-12}$ 	  & Co 	K$_{\beta1}$, Ni K$_{\alpha1}$, K$_{\alpha2}$	&-						\\
 	&         		&		&				&	      			  &                                     & 	   	   						\\ 
\hline
 4U 1538$-$522	&1.26$^{+0.05}_{-0.07}$	& E	& 0.22$^{+0.07}_{-0.04}$  	& 321$^{+102}_{-58}$	  & Mg K$_{\alpha1}$, K$_{\alpha2}$, K$_{\beta1}$ 	& 1.34 (fixed), Mg K$_{\alpha}$, Mg XI He$_{\alpha}$ (eclipse) $^{[y1]}$	\\
 (0152780201)	& 			&	&	      			&    	   	   		  &                                                     &				\\ 
 	        &1.85$^{+0.03}_{-0.03}$	& E	& 0.15$^{+0.02}_{-0.02}$ 	& 250$^{+33}_{-33}$ 	  & Si K$_{\beta1}$ 					&1.848$^{+0.012}_{−0.024}$, Si XIII He$_{\alpha}$ (eclipse) $^{[y1]}$		\\
		& 			&	&	      			&    	   	   		  &				                        &                                                                             \\ 
		&2.40$^{+0.04}_{-0.03}$	& E	& 0.07$^{+0.01}_{-0.01}$  	& 116$^{+17}_{-17}$	  & S K$_{\beta1}$ 					&2.4427$^{+0.0023}_{−0.023}$, S XV He$_{\alpha}$ (eclipse) $^{[y1]}$	\\
		& 			&	&	      			&                                 &                                                     & 	   	   						\\ 
		&6.02$^{+0.05}_{-0.05}$	& E	& 0.04$^{+0.01}_{-0.01}$ 	& 54$^{+14}_{-14}$	  & Cr K$_{\beta1}$					& -		\\
		&			&	&                               &    	   	   	          &					\\ 
\hline
IGR J17252$-$3616&7.01$^{+0.14}_{-0.40}$& E & 0.05$^{+0.04}_{-0.01}$  	& 1227$^{+982}_{-245}$	  & 	-	                                        &	-	\\
(0405640201)   	 &		        &   &				&               	   	   						\\ 

\hline
 \end{tabular}
 \label{elements_alltogether}
  \footnotesize{${v1}$: \citep{1996Ebisawa}}
  \footnotesize{${v2}$: \citep{2001Wojdowski}}
  \footnotesize{${w1}$: \citep{2009Neilsen}}
  \footnotesize{${x1}$: \citep{2005van_der_Meer}}
  \footnotesize{${y1}$: \citep{2011Roca}}
  \end{table*}
%
\begin{figure*}
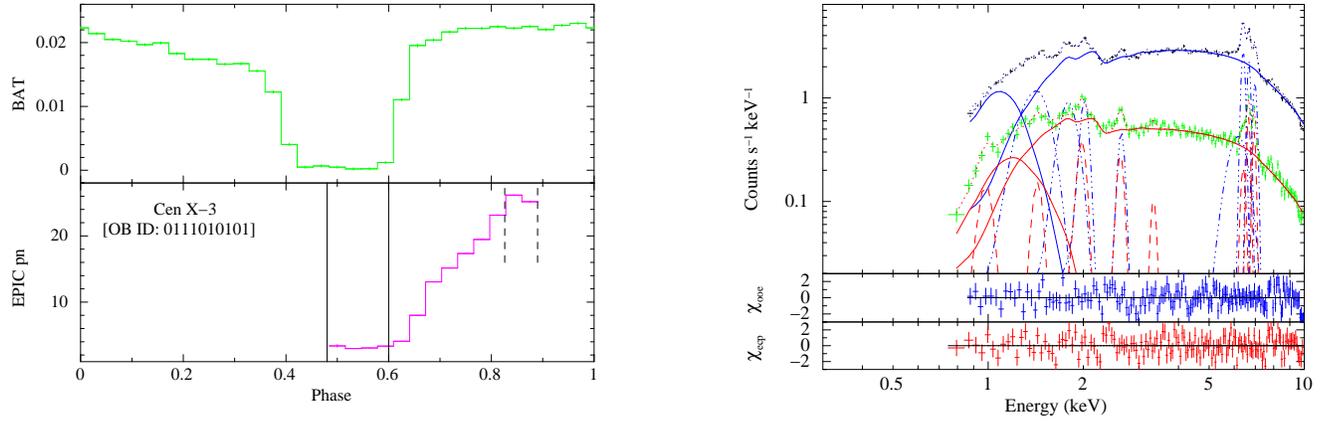

\centering
\begin{subfigure}
\centering
\includegraphics[scale=0.32, angle=-90, trim={0 0 0 0}, clip]{efold_BAT_XMM_final_cenx3.ps}
\end{subfigure}
\hspace{1 cm}
\begin{subfigure}
\centering
\includegraphics[scale=0.3, angle=-90, trim={0 0 0 0}, clip]{same_E-range_cenx3.ps}
\end{subfigure}
\caption[Cen X-3 orbital profile, eclipse and out-of-eclipse spectra]{\textbf{Left figure:} 
Top panel shows the long term average BAT orbital profile of  Cen X-3 
and the bottom panel shows  EPIC pn light curve of 
Cen X-3 (OB ID: 0111010101) folded with its orbital period. The eclipse spectrum and the 
out-of-eclipse spectrum was extracted from the duration shown
with the solid and the dashed lines respectively. 
\textbf{Right figure:} Top Panel gives the EPIC pn out-of-eclipse and the eclipse spectrum of Cen X-3, 
where datapoints for the out-of-eclipse spectrum are plotted
with colour black and that of eclipse spectrum are plotted with colour green. Model components of 
the out-of-eclipse and the eclipse spectrum is 
shown with the colour blue and  red respectively. The emission line for the out-of-eclipse spectrum
is shown with blue lines and those for the eclipse spectrum are shown with red lines.
The middle and bottom panels show the
contribution of each bin towards $\chi$ for
 the best fit spectral model for the out-of-eclipse spectrum and the eclipse spectrum respectively.}
\label{fig:cenx3}
\end{figure*}

\begin{figure*}
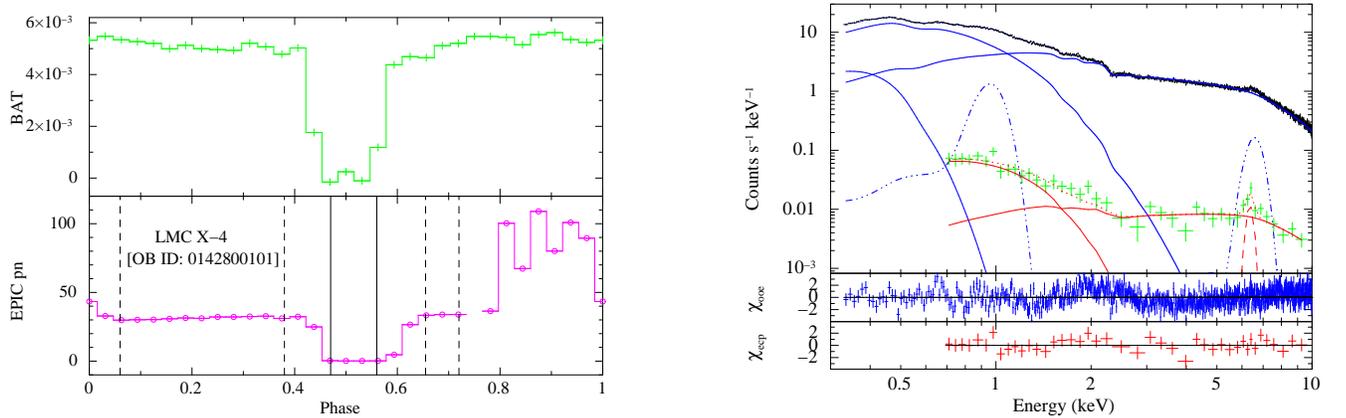

\centering
\begin{subfigure}
\centering
\includegraphics[scale=0.32, angle=-90]{efld_BAT_XMM_LMCX4_id101.ps}
\end{subfigure}
\hspace{1 cm}
\begin{subfigure}
\centering
\includegraphics[scale=0.3, angle=-90]{same_E-range_LMCX4_ID101_26june18_1.ps}
  \end{subfigure}
 \caption{Same as Figure \ref{fig:cenx3} for LMC X-4 (OB ID: 0142800101)}
\label{fig:lmcx4_101}
\end{figure*}

\begin{figure*}
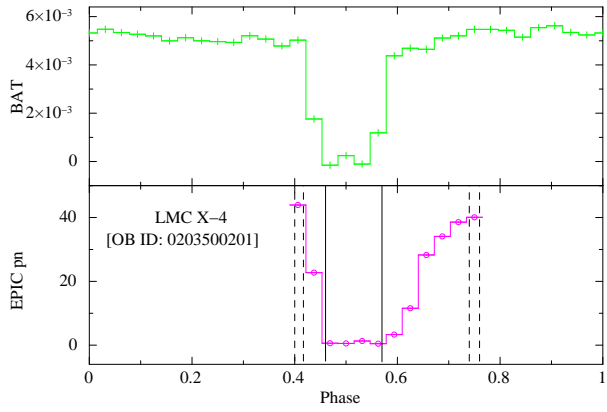

\centering
\begin{subfigure}
\centering
\includegraphics[scale=0.32, angle=-90]{efld_BAT_XMM_LMCX4_id201_TM-mode.ps}
\end{subfigure}
\hspace{1 cm}
\begin{subfigure}
\centering
\includegraphics[scale=0.3, angle=-90]{same_E-range_LMCX4_ID201_26june18_1.ps}
\end{subfigure}
\caption{Same as Figure \ref{fig:cenx3} for LMC X-4 (OB ID: 0203500201)}
\label{fig:lmcx4_201TM}
\end{figure*}

\begin{figure*}
\centering
\begin{subfigure}
\centering
\includegraphics[scale=0.32, angle=-90]{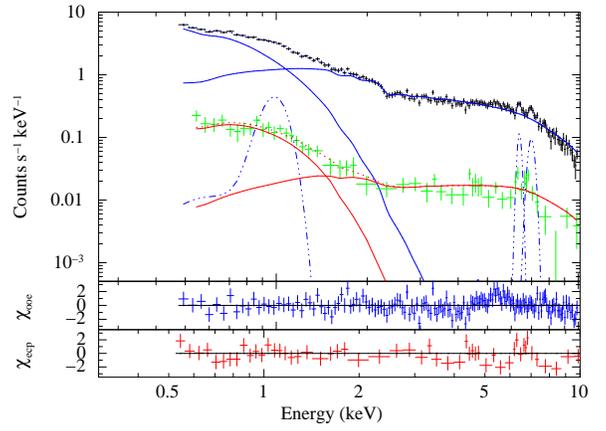}
\end{subfigure}
\hspace{1 cm}
\begin{subfigure}
\centering
\includegraphics[scale=0.3, angle=-90]{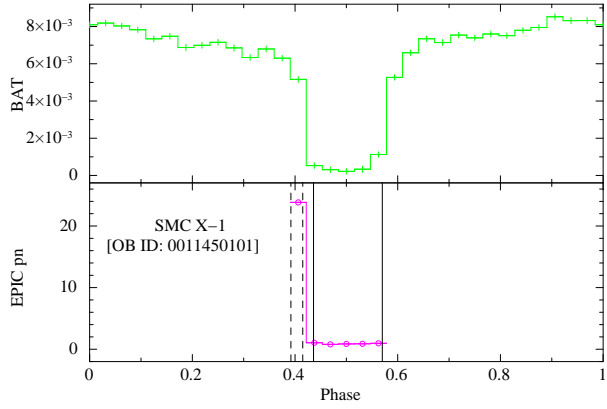}
\end{subfigure}
\caption{Same as Figure \ref{fig:cenx3} for SMC X-1 (OB ID: 0011450101)}
\label{fig:smcx1}
\end{figure*}

\begin{figure*}
\centering
\begin{subfigure}
\centering
\includegraphics[scale=0.32, angle=-90]{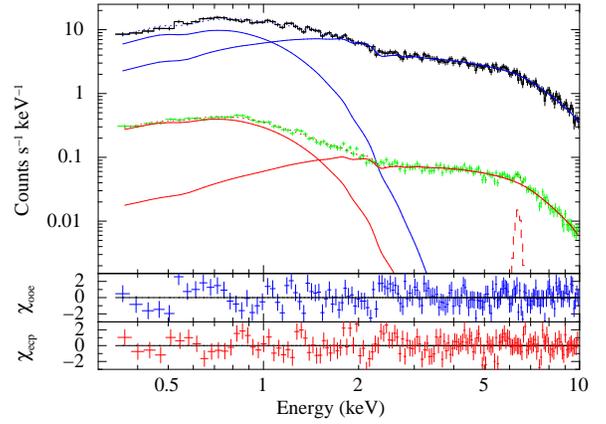}
\end{subfigure}
\hspace{1 cm}
\begin{subfigure}
\centering
\includegraphics[scale=0.32, angle=-90]{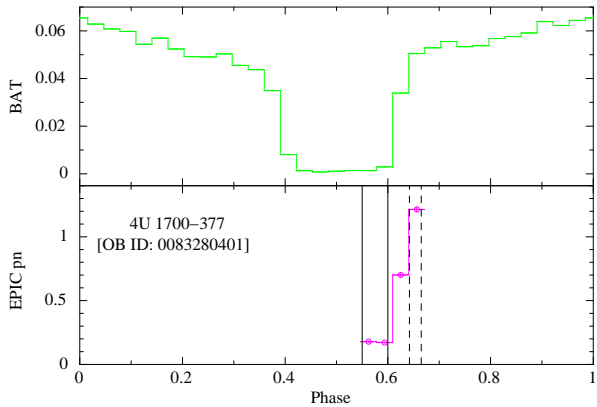}
\end{subfigure}
\caption{Same as Figure \ref{fig:cenx3} for 4U 1700$-$377 (OB ID: 0083280401)}
\label{fig:U1700_id401}
\end{figure*}



\begin{figure*}
\centering
\begin{subfigure}
\centering
\includegraphics[scale=0.32, angle=-90]{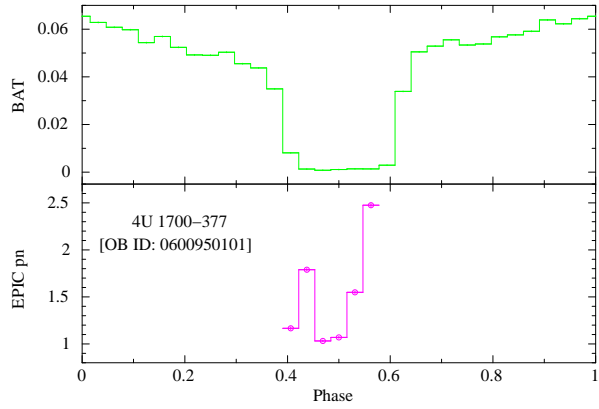}
\end{subfigure}
\hspace{1 cm}
\begin{subfigure}
\centering
\includegraphics[scale=0.3, angle=-90, trim={0 0 0 0}, clip]{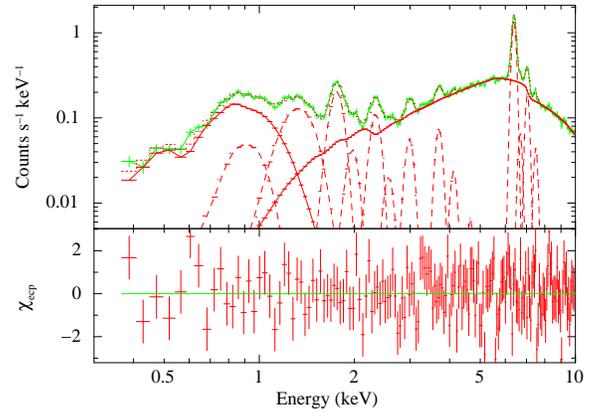}
\end{subfigure}
\caption[4U 1700$-$377 (OB ID: 0600950101) orbital profile and eclipse spectrum]{\textbf{Left figure:} 
Top panel shows the long term average BAT orbital profile of 4U 1700$-$377  and the bottom panel shows  EPIC pn light curve of 
 4U 1700$-$377 (OB ID: 0600950101) folded with its orbital period. 
\textbf{Right figure:} Top Panel gives the EPIC pn eclipse spectrum of 4U 1700$-$377 (OB ID: 0600950101).
The bottom panels show the residuals to the best fit spectral model for the eclipse spectrum.}
\label{fig:spec_1700_ecp}
\end{figure*}

\begin{figure*}
\centering
\begin{subfigure}
\centering
\includegraphics[scale=0.32, angle=-90]{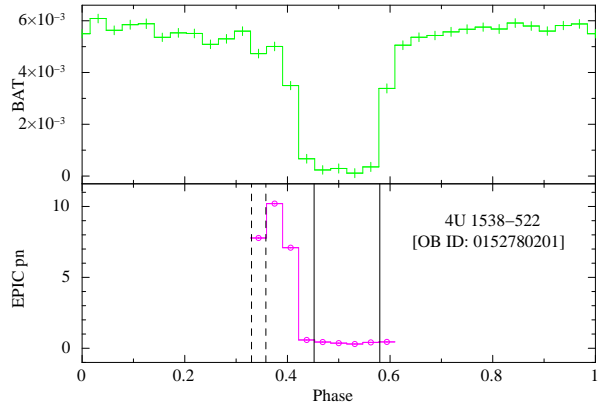}
\end{subfigure}
\hspace{1 cm}
\begin{subfigure}
\centering
\includegraphics[scale=0.3, angle=-90]{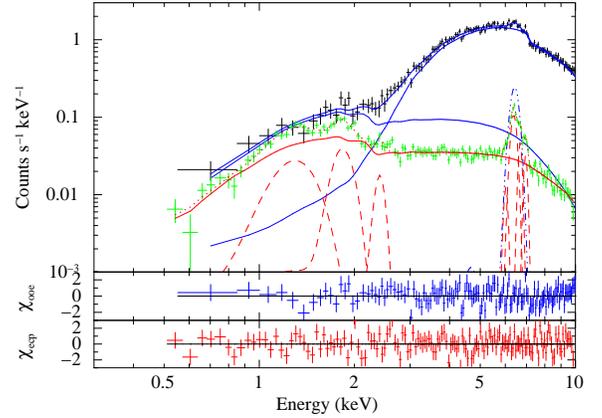}
\end{subfigure}
\caption{Same as Figure \ref{fig:cenx3} for 4U 1538$-$522 (OB ID: 0152780201)}
\label{fig:U1538}
\end{figure*}

\begin{figure*}
\centering
\begin{subfigure}
\centering
 \includegraphics[scale=0.32, angle=-90]{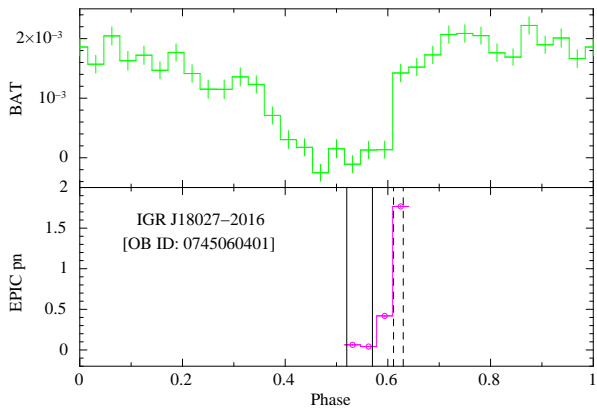}
\end{subfigure}
\hspace{1 cm}
\begin{subfigure}
\centering
\includegraphics[scale=0.32, angle=-90]{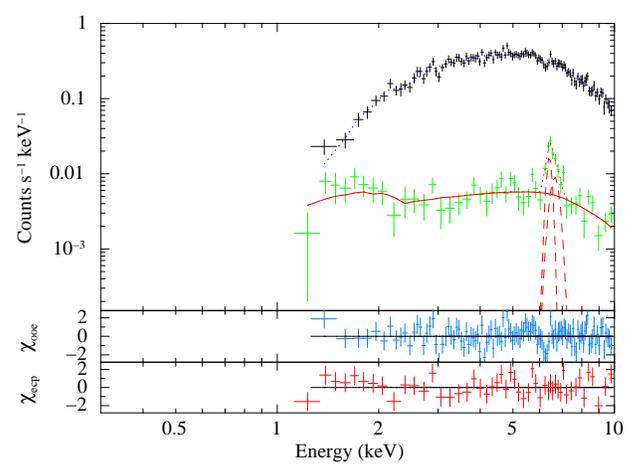}
\end{subfigure}
\caption{Same as Figure \ref{fig:cenx3} for IGR J18027$-$2016 (OB ID: 0745060401)}
\label{fig:IGR2016}
\end{figure*}
%
\begin{figure*}
\centering
\begin{subfigure}
\centering
\includegraphics[scale=0.32, angle=-90]{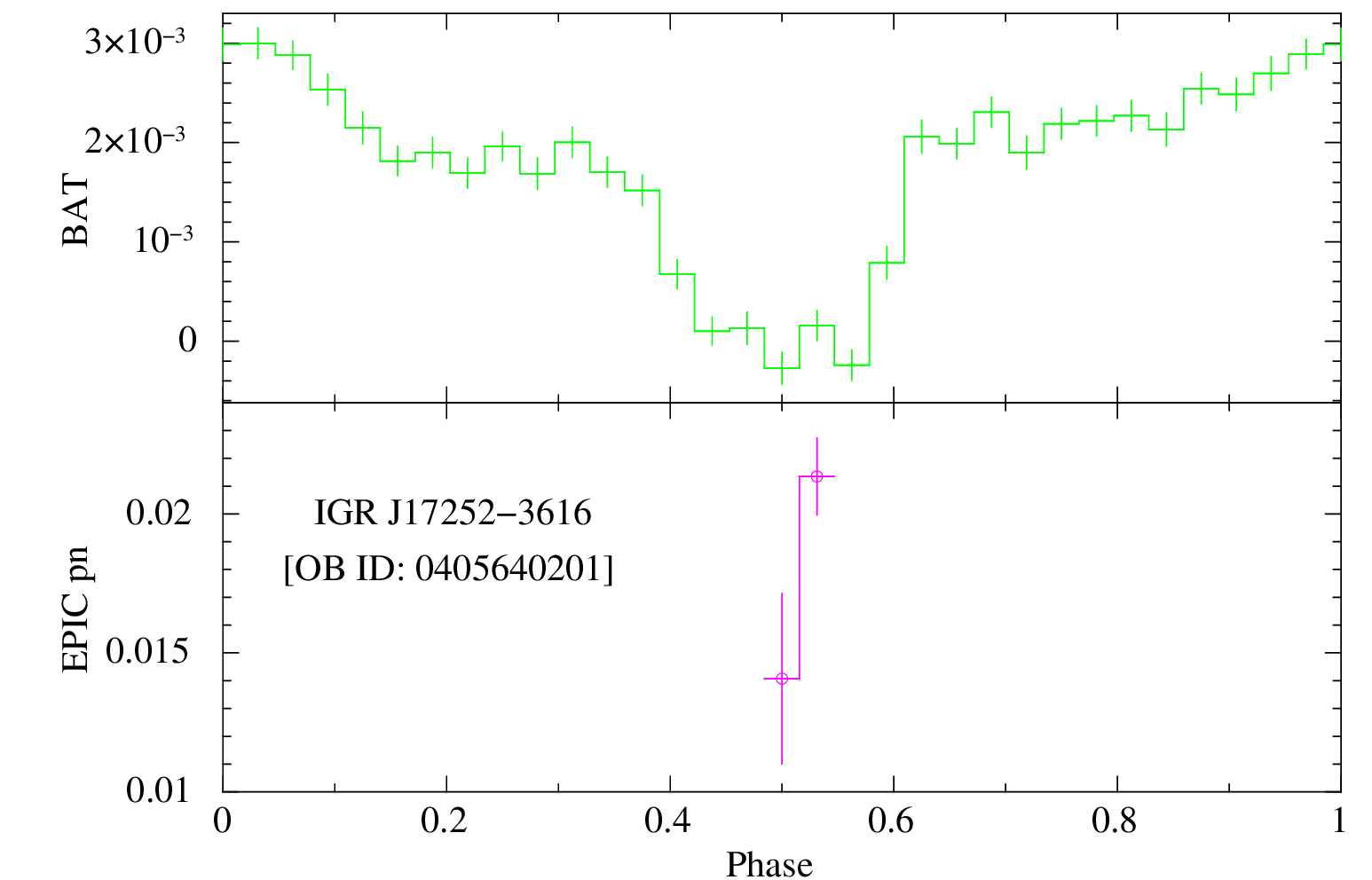}
\end{subfigure}
  \hspace{1 cm}
\begin{subfigure}
\centering
\includegraphics[scale=0.3, angle=-90]{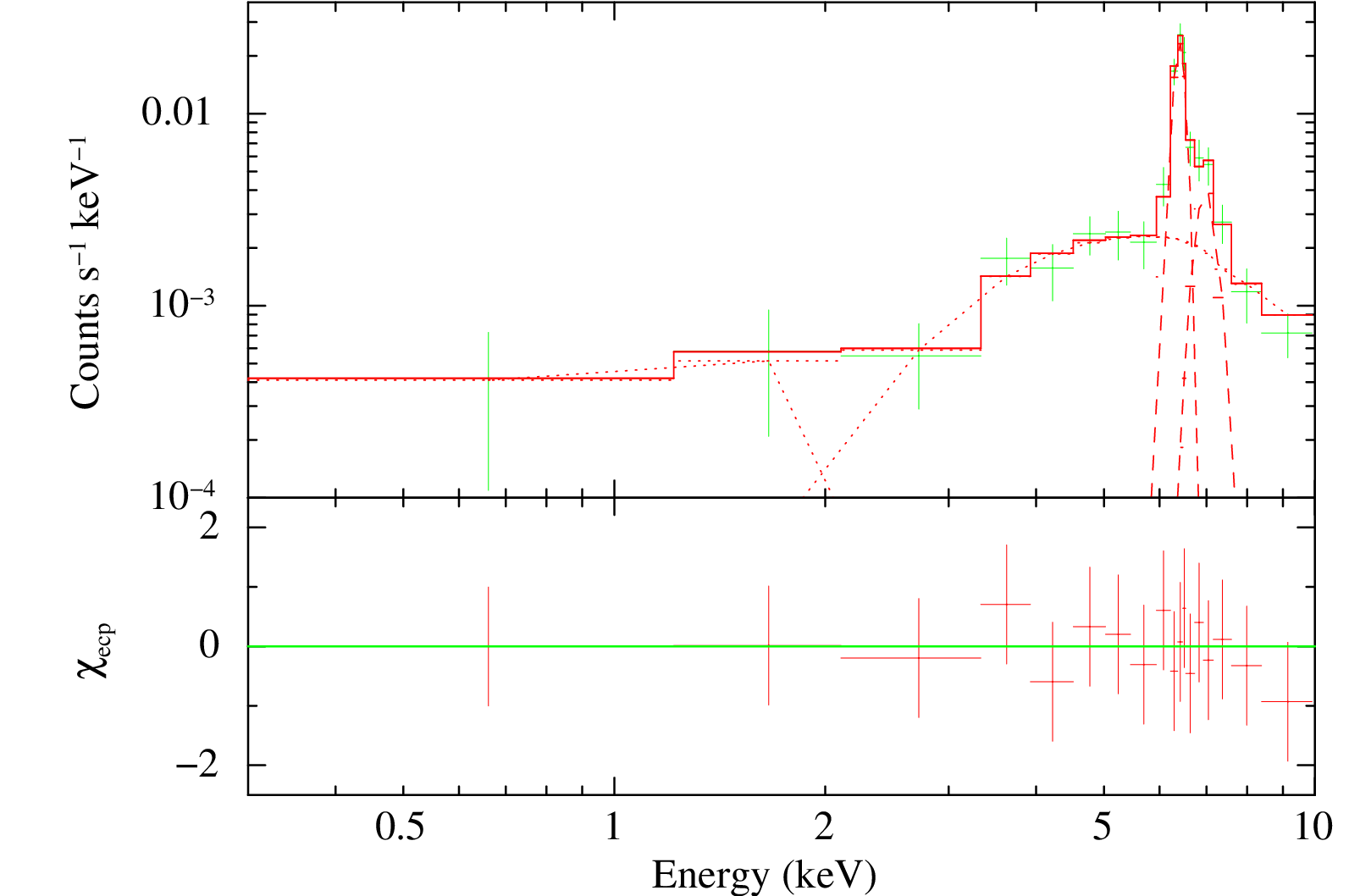}
\end{subfigure}
\caption{Same as Figure \ref{fig:spec_1700_ecp} for IGR J17252$-$3616 (OB ID: 0405640201)}
\label{fig:IGR3616_201}
\end{figure*}
%
\begin{figure*}
\centering
\begin{subfigure}
\centering
\includegraphics[scale=0.3, angle=-90]{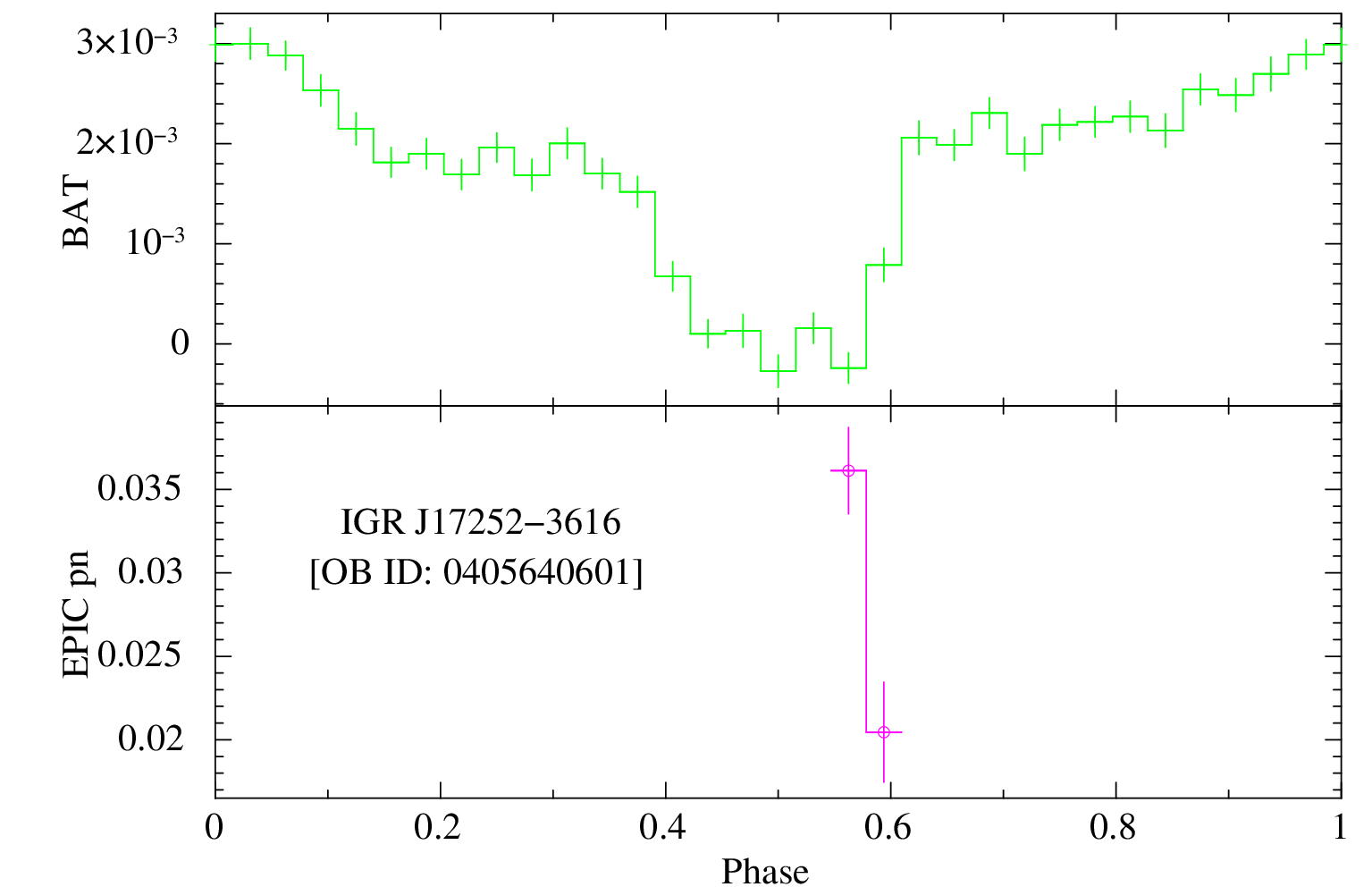}
\end{subfigure}
\hspace{1 cm}
\begin{subfigure}
\centering
\includegraphics[scale=0.3, angle=-90]{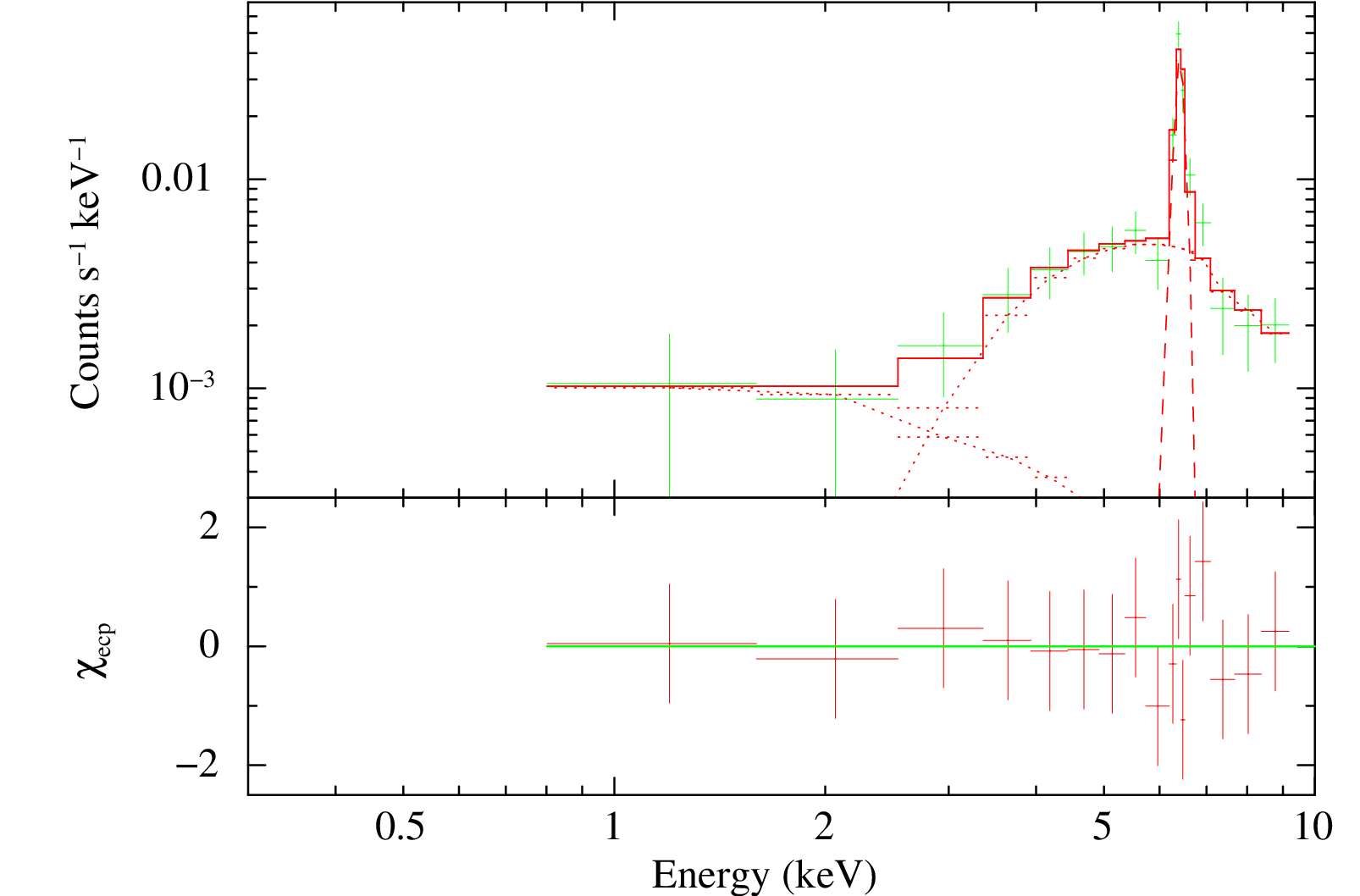}
\end{subfigure}
\caption{Same as Figure \ref{fig:spec_1700_ecp} for IGR J17252$-$3616 (OB ID: 0405640601)}
\label{fig:IGR3616_601}
\end{figure*}
%
\begin{figure*}
\centering
\begin{subfigure}
\centering
\includegraphics[scale=0.3, angle=-90]{efld_BAT_XMM_IGRJ17252-3616_id1001.ps}
\end{subfigure}
\hspace{1 cm}
\begin{subfigure}
\centering
\includegraphics[scale=0.3, angle=-90]{same_E-range_IGRJ17252-3616_id1001_1.ps}
\end{subfigure}
\caption{Same as Figure \ref{fig:spec_1700_ecp} for IGR J17252$-$3616 (OB ID: 0405641001)}
\label{fig:IGR3616_1001}
\end{figure*}
\begin{figure*}
\centering
\begin{subfigure}
\centering
\includegraphics[scale=0.3, angle=-90]{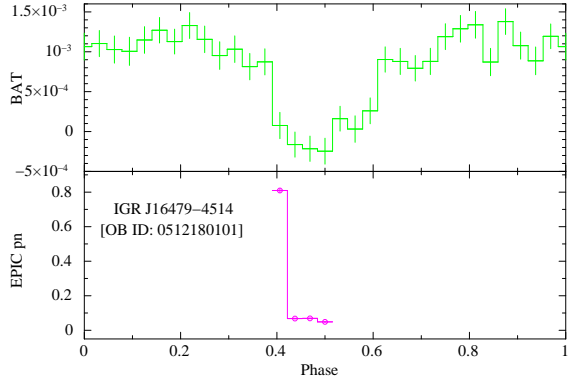}
\end{subfigure}
\hspace{1 cm}
\begin{subfigure}
\centering
\includegraphics[scale=0.3, angle=-90]{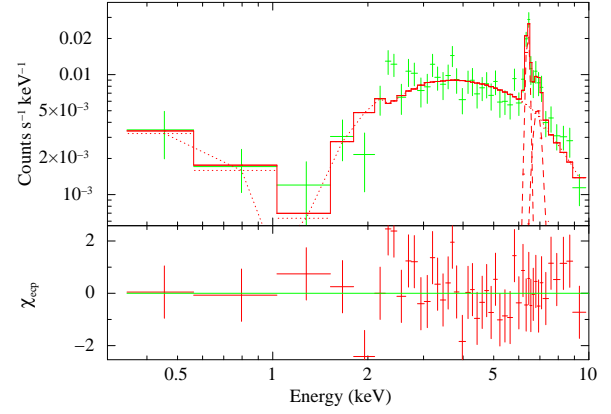}
\end{subfigure}
\caption{Same as Figure \ref{fig:spec_1700_ecp} for IGR J16479$-$4514 (OB ID: 0512180101)}
\label{fig:IGR4514}
\end{figure*}
%
\begin{figure*}
\centering
\begin{subfigure}
\centering
\includegraphics[scale=0.3, angle=-90]{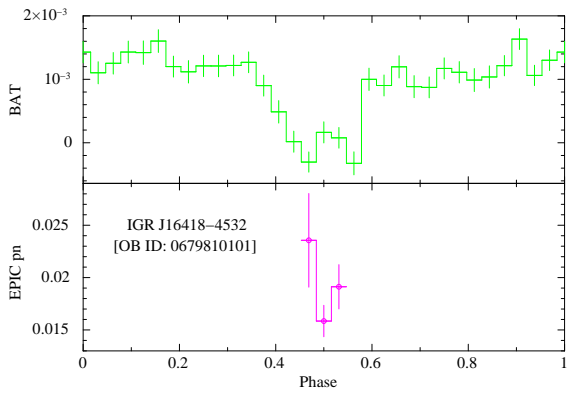}
\end{subfigure}
\hspace{1 cm}
\begin{subfigure}
\centering
\includegraphics[scale=0.3, angle=-90]{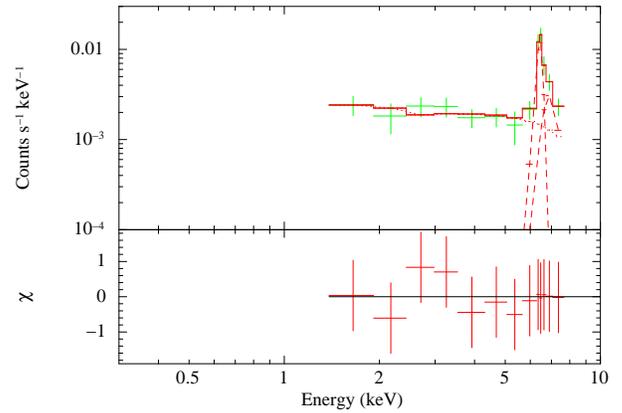}
\end{subfigure}
\caption{Same as Figure \ref{fig:spec_1700_ecp} for IGR J16418$-$4532 (OB ID: 0679810101)}
\label{fig:IGR4532}
\end{figure*}
%
%
\clearpage

\begin{figure*}
\centering
 \includegraphics[scale=0.6, angle=-90]{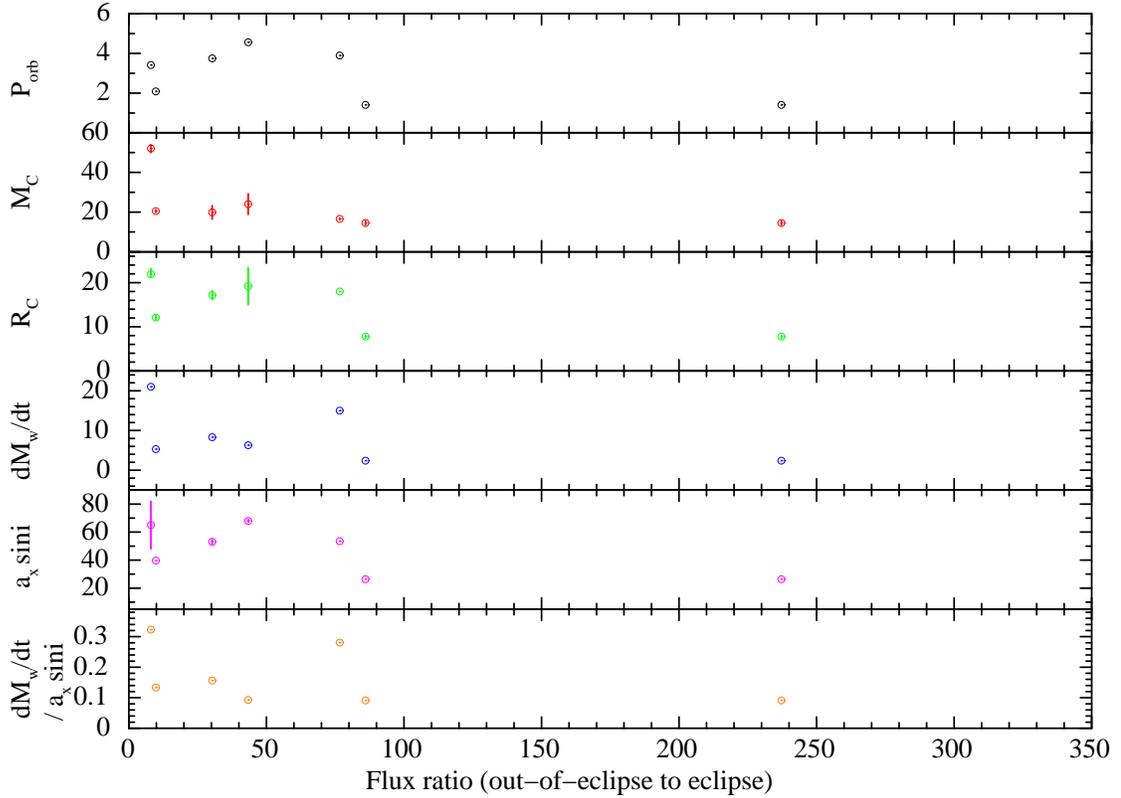}
 \caption{Plot of different orbital parameters of the eclipsing HMXBs with out-of-eclipse to eclipse flux ratios.
 P$_{orb}$: Orbital Period (days), M$_{\rm C}$: Mass of the companion (M$_{\odot}$), R$_{\rm C}$: Radius of the
 companion star (R$_{\odot}$), $\dot{\rm M}_{\rm w}$: Mass loss rate of the companion star (10$^{-7}~ \rm M_{\odot} \rm ~yr^{-1}$),
 a$_{x}$sini: Projected length of the semi major axis of the system in the plane along the line of sight (light-sec).
 M$_{\odot}$, R$_{\odot}$: Mass and radius of Sun respectively.}
 \label{6param_flux-ratio}
\end{figure*}

\begin{figure*}
\centering
 \includegraphics[scale=0.4, angle=-90]{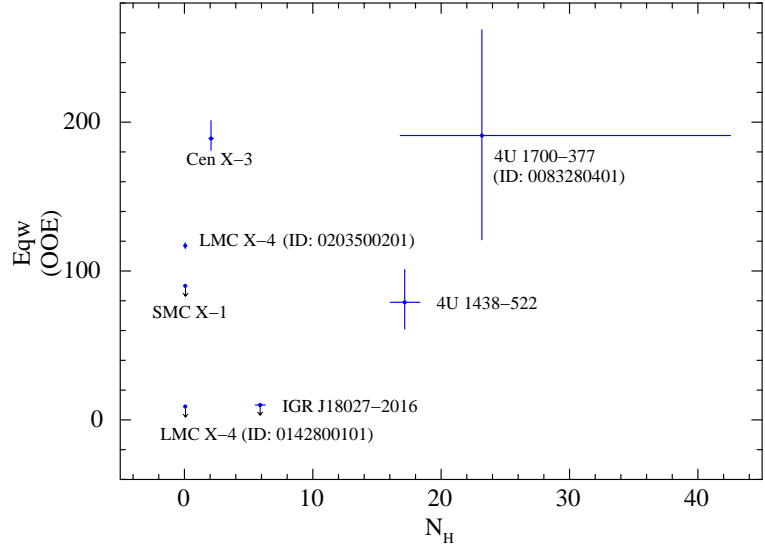}
 \caption{Equivalent width (Eqw, along Y axis) vs. line of sight equivalent Hydrogen column density (N$_{\rm H}$, along X axis) 
 of fluorescent Fe K$_{\alpha}$ emission line in out-of-eclipse (OOE) phase. This line is detected in Cen X-3, 
 LMC X-4 (ID: 0203500201), 4U 1700$-$477 (ID: 0083280401) and in 4U 1538$-$522, whereas upper limit of equivalent width is shown in
 SMC X-1, LMC X-4 (ID: 0142800101) and IGR J18027$-$2016. Unfortunately one can not correlate the relationship between N$_{\rm H}$
 and equivalent width of fluorescent Fe K$_{\alpha}$ emission line in this sample of SgHMXBs due to the limitation of the available data.}
 \label{OOE_NH-eqw}
\end{figure*}

%
\begin{figure*}
\centering
 \includegraphics[scale=0.5, angle=-90]{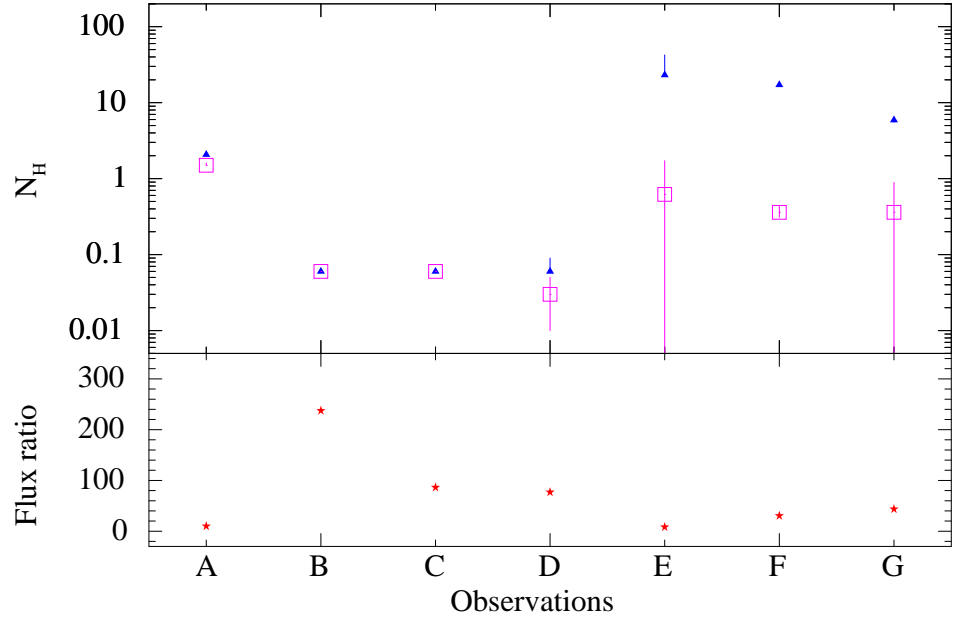}
  \caption{Line of sight equivalent Hydrogen column
 density (N$_{\rm H}$) of HMXBs during eclipse and out-of-eclipse
 phase for observations having data for both the phases. X axis gives the observations and the Y 
 axis in the top panel shows N$_{\rm H}$ (eclipse: pink boxes, out-of-eclipse: blue triangles).
 Y axis in the bottom panel gives the value of
 0.3-10 keV out-of-eclipse to eclipse flux ratio. 
 A: Cen X-3, B: LMC X-4 OB ID: 0142800101, C: LMC X-4 OB ID: 0203500201,
 D: SMC X-1, E: 4U 1700$-$377 OB ID: 0083280401, F: 4U 1538$-$522, G: IGR J18027$-$2016.
 For 3 observations (4U 1700$-$377 OB ID: 0083280401,  4U 1538$-$522, IGR J18027$-$2016)
 N$_{\rm H}$ is larger during eclipse  than out-of-eclipse   
 phase. For the other4 observations N$_{\rm H}$ is comparable in both the phases.}
 \label{NH_flux}
\end{figure*}
\begin{figure*}
\centering
\includegraphics[scale=0.5, angle=-90]{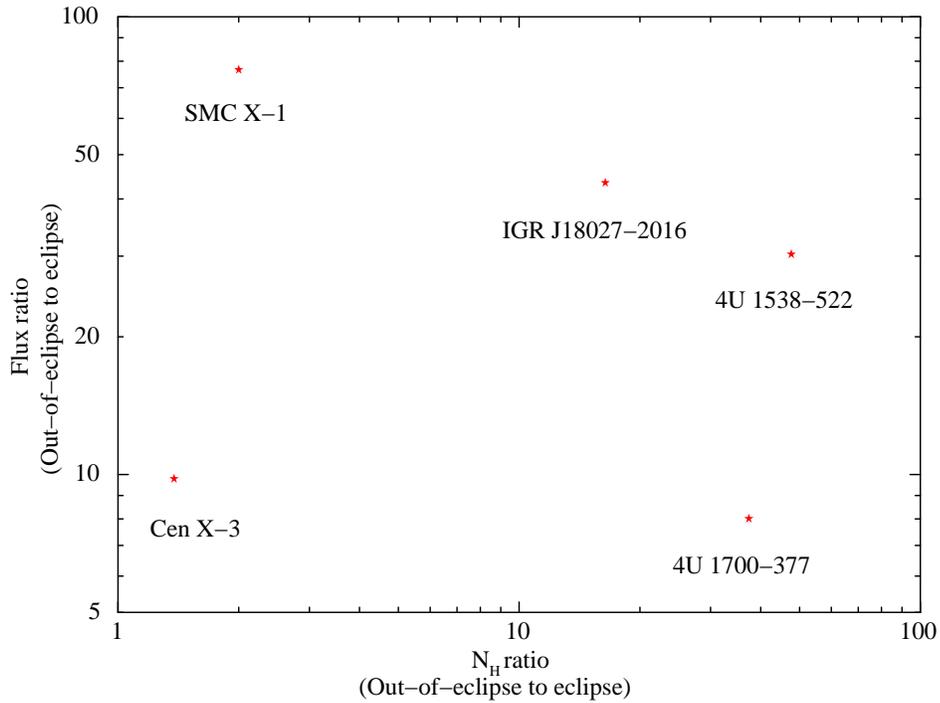}
\caption{Variation of out-of-eclipse to eclipse flux ratio with corresponding ratio of the line of sight equivalent Hydrogen column
 density (N$_{\rm H}$) of HMXBs. X axis shows the out-of-eclipse to eclipse ratio of N$_{\rm H}$ and Y axis gives the 
 out-of-eclipse to eclipse flux ratio. Here the observation of 4U 1700$-$377 belongs to OB ID: 0083280401}
 \label{NH_flux_ratio_relation}
\end{figure*}
%
\begin{figure*}
\vspace{-0.45cm}
\centering
  \includegraphics[scale=0.22, angle=0]{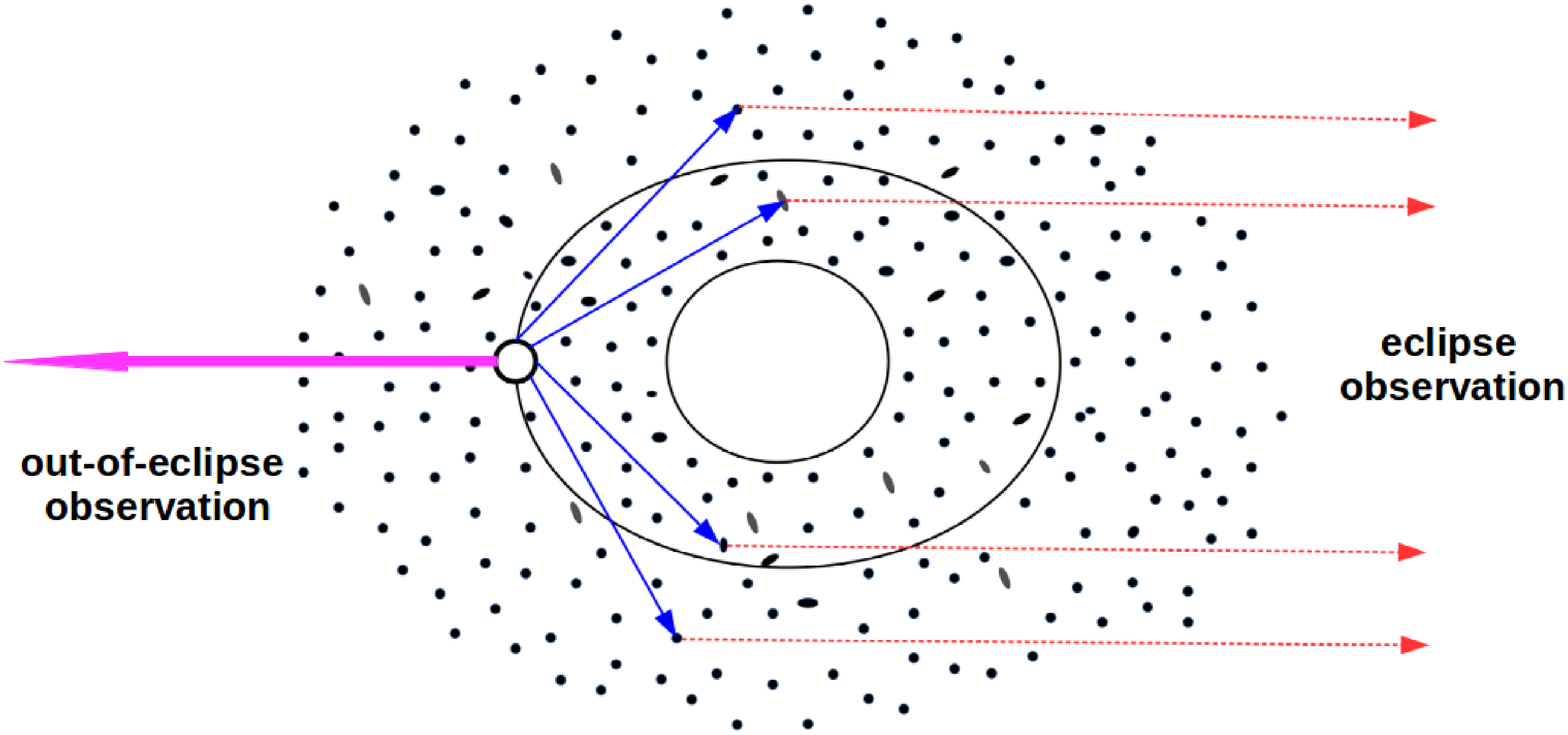}
  \hspace{1cm}
  \includegraphics[scale=0.22, angle=0]{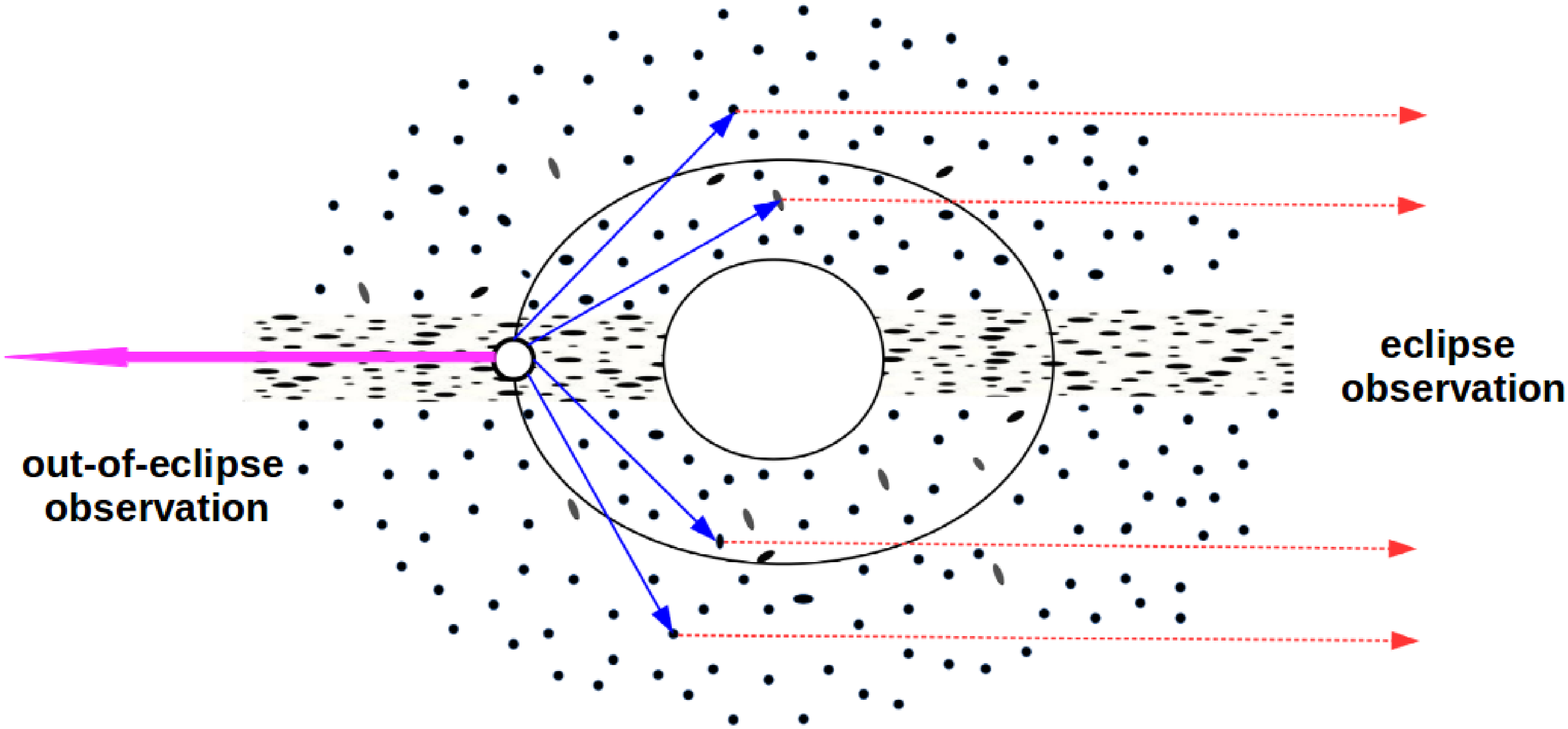}  
 \caption{A probable distribution of wind around the neutron star in systems with low ratio of the 
 column density (N$_{\rm H}$) in the out-of-eclipse to eclipse phase (left) and in systems in which this ratio of 
 column densities is much larger (right). The small circle at the left of each Figure 
 represents the neutron star and the bigger one represents the companion star. 
 The primary and reprocessed X-rays are represented with solid and dashed arrows respectively. 
  The first system probably
 have more isotropic wind pattern. A strong wind outflow in the equatorial or near equatorial plane
 in the second system can explain the higher ratio of column densities between the out-of-eclipse to eclipse phase.
 During eclipse, the reprocessed X-rays represented with the dashed arrows reach the observer and in both the systems it 
 shows comparable N$_{\rm H}$. When the source is in out-of-eclipse phase the observer receives the direct X-rays, shown here
 with bold (pink) arrows. In case of the first systems during out-of-eclipse phase, the observer looks through an isotropic wind  
 as seen during eclipse. But for the second system during out-of-eclipse phase,
 the direct X-rays reach the observer penetrating dense wind.}
 \label{dense_flow}
\end{figure*}
%


 \end{document}